\documentclass[aps,twocolumn,groupedaddress,floatfix]{revtex4}
\usepackage[colorlinks=true]{hyperref} 
\hypersetup{urlcolor=blue,linkcolor=red,citecolor=blue,colorlinks=true}

\usepackage{graphics,epsfig}
\usepackage{amssymb,amsfonts}
\usepackage{amsmath}
\usepackage{mathtools}
\usepackage{color}
\usepackage{tikz}

\usepackage[export]{adjustbox}

\newcommand{\bb}[0]{\begin{eqnarray}}
\newcommand{\ee}[0]{\end{eqnarray}}

\newcommand{\eref}[1]{Eq.~(\ref{#1})}
\newcommand{\efig}[1]{Fig.~\ref{#1}}

\newcommand{\bw}{\begin{widetext}}
\newcommand{\ew}{\end{widetext}}

\begin{document}
\title{Quantum consistent neural/tensor networks for photonic circuits with strongly/weakly entangled states.}

\author{Nicolas Allegra}
\email{nicolas.allegra@gmail.com}
\begin{abstract}
Modern quantum optical systems such as photonic quantum computers and quantum imaging devices require great precision in their designs and implementations in the hope to realistically exploit entanglement and reach a real quantum advantage. The theoretical and experimental explorations and validations of these systems are greatly dependent on the precision of our classical simulations. However, as Hilbert spaces increases, traditional computational methods used to design and optimize these systems encounter hard limitations due to the quantum curse of dimensionally. To address this challenge, we propose an approach based on neural and tensor networks to approximate the exact unitary evolution of closed entangled systems in a precise, efficient and quantum consistent manner. By training the networks with a reasonably small number of examples of quantum dynamics, we enable efficient parameter estimation in larger Hilbert spaces, offering an interesting solution for a great deal of quantum metrology problems.
 \end{abstract}

\date{\today}

\maketitle
\section{Introduction}
Quantum metrology and parameter estimation have emerged as forefront research areas, pushing the boundaries of precision measurements beyond classical limits \cite{giovannetti2011advances,toth2014quantum,polino2020photonic,barbieri2022optical}. Capitalizing on the nature of quantum physics might open great opportunities to enhance measurement precision and sensitivity, with potential applications ranging from fundamental physics \cite{schnabel2010quantum} to cutting-edge technologies \cite{taylor2016quantum,dowling2014quantum,sidhu2021advances}. In the quantum world, the Heisenberg uncertainty principle sets a fundamental limit on the precision with which certain pairs of observables, such as position and momentum, can be simultaneously measured. Quantum metrology offers to push these classical bounds by taking advantage of entanglement properties, enabling measurements with precision beyond what is achievable in classical systems.

 One of the key components of quantum metrology is quantum estimation theory \cite{helstrom1969quantum,toth2014quantum}, which provides a rigorous framework for optimizing measurement strategies to extract the maximum amount of information from quantum states. This involves harnessing quantum resources, to minimize uncertainties and enhance the precision of parameter estimation. Quantum estimation techniques are crucial for designing experiments that leverage quantum advantages, ensuring optimal use of quantum resources for a given measurement task. This promises transformative advances in fields such as gravitational wave detection\cite{saulson1994fundamentals} , quantum imaging \cite{kolobov2007quantum} and communication \cite{gisin2007quantum}, quantum computing \cite{nielsen2010quantum} and quantum machine learning \cite{biamonte2017quantum}.
In recent years, significant progress has been made in the intersection of deep learning and quantum mechanics; numerous studies have demonstrated effective techniques for learning quantum dynamics, showcasing applications in diverse areas such as condensed matter systems, quantum optics, and quantum computation \cite{carrasquilla2020machine,gebhart2023learning,torlai2018neural,ahmed2021classification,ahmed2021quantum,cimini2023deep,lange2024architectures}. These new techniques combined with powerful theoretical and numerical quantum methods, such as matrix product states and tensor networks, could greatly improve classical simulations of complex quantum systems as well as finding potential solutions to previously insurmountable problems (see \cite{dawid2022modern} for a recent review).  

In particular, the field of linear optical quantum computing, which is a paradigm of quantum computation allowing universal quantum computation using photons as information carriers, linear optical elements as computation elements, and photon detectors and quantum memories to detect and store quantum information \cite{divincenzo2000physical,knill2001scheme,nielsen2004optical,kok2007linear,o2007optical,kok2010introduction}, is especially impacted by the rapid convergence between high-precision integrated photonic systems, sophisticated algorithmic protocols and modern machine learning. Very recently, optical quantum computers based on photonic circuits or integrated photonic chips, built by academic institutions \cite{zhong2020quantum,zhong2021phase} and private companies \cite{arrazola2021quantum,madsen2022quantum,maring2024versatile} such as Xanadu in Canada or Quandela in France have reached very high fidelities and showed great potential in the race for the first useful quantum computer.  The advantages of high coherence, high processing speed and reasonably high working temperature are offset by the precision requirements for high fidelity quantum computation which could be greatly enhance, thanks to the ever-growing advances in deep learning, in the perspective of building commercially available and programmable quantum computers.

In the end of their very insightful and inspiring textbook \cite{kok2010introduction}, Pieter Kok and Brendon W. Lovett wrote "It seems that large-scale quantum metrology is as hard to implement as large scale quantum computing". By putting the two topics in a single book, they suggest that optical quantum metrology and optical quantum computing are two sides of the same coin. We shall explore this relationship throughout the paper by studying classical simulations of quantum metrology problems in the context of optical quantum computers.

\newpage

\textbf{Outline of the paper:} In the first section of this paper, we briefly describe the basic concepts governing the unitary evolution of closed and entangled quantum systems that are relevant for the ever-growing field of photonic quantum computers. Next, we formulate a simple and relevant problem of quantum parameter estimation, and show how the versatile tools of automatic differentiation and back-propagation can be used to address it. The discussion will then focus on the limitations inherent in this approach as the Hilbert space grows larger and on a reformulation of the problem through the lens of modern classical simulations and deep learning. By employing parameterized networks that respect the invariances and symmetries of the quantum system as well as operating within a significantly reduced space, we show how to approximate the exact unitary evolution precisely in different cases of entanglement strength. Upon successful training of this network, back-propagated quantum parameter estimation algorithms can be readily applied, offering an efficient and scalable solution to real systems. Finally we will explore how to generalize the approach for more complex cases and discuss the difficulties, as well as potential avenues to solve them.

\section{Random unitary evolution and entanglement}
\begin{figure}[ht!]
\includegraphics[scale=0.15]{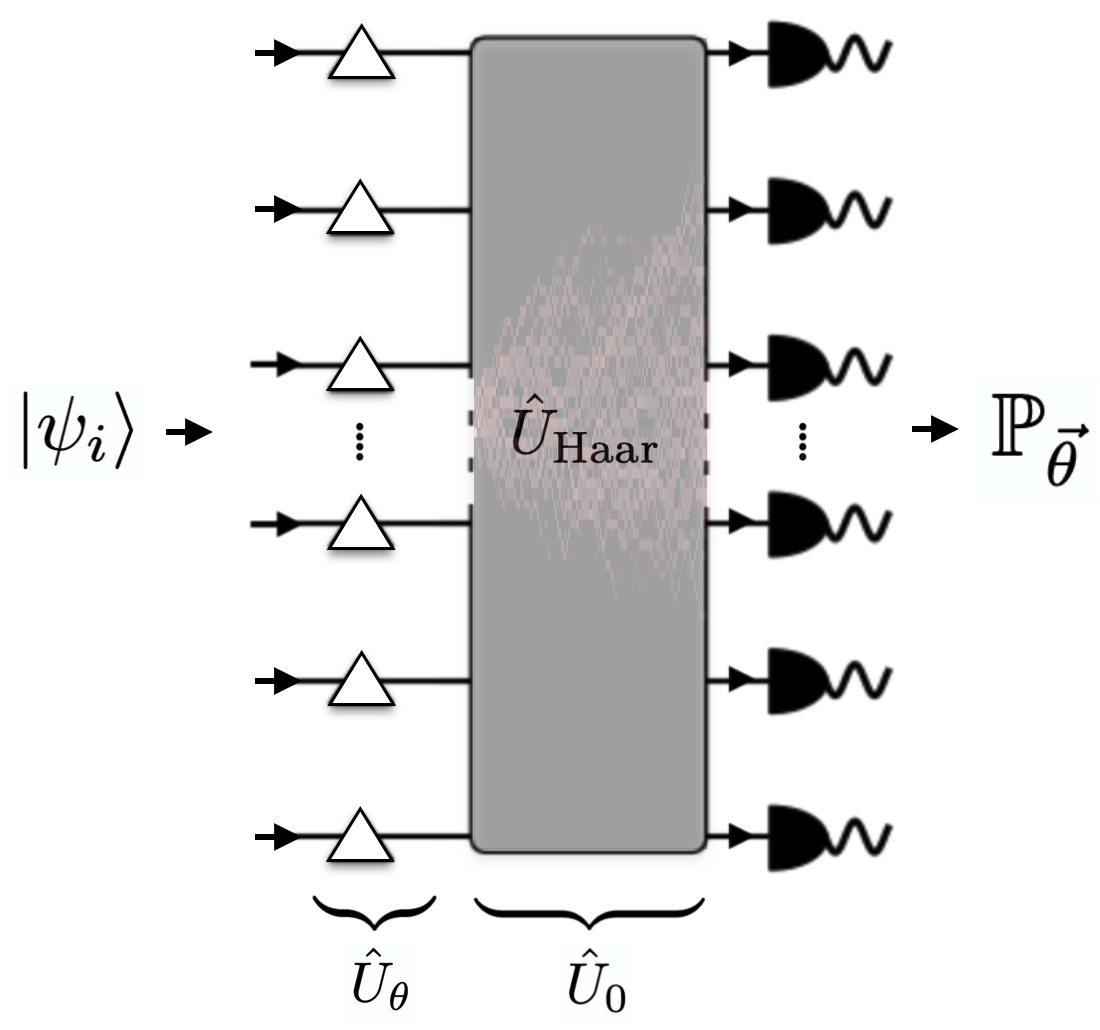} 
	\caption{Illustration of the random unitary circuit with phase-shifter (triangles) and photon counting detectors (semi-circles)}
	\label{fig_circuit}
\end{figure}

In order to account for systems that are ubiquitous in quantum optical computation and quantum information science, let us consider a system like the one sketched in \efig{fig_circuit}.
It consists in the Heisenberg evolution of an initial bosonic state $\vert\psi_i\rangle$ described by
\begin{equation}\label{heisenberg}
\vert \psi_{\vec\theta}\rangle = \mathrm{Sym}\  \hat{U}\vert \psi_i \rangle,
\end{equation}
where $\hat{U}=\hat{U}_{0}.\hat{U}_{\vec\theta}$ is the unitary evolution operator describing the multi-mode bosonic system and decomposed into a non-controllable component $\hat{U}_{0}$ and a parameterized component $\hat{U}_{\vec\theta}$ and $\mathrm{Sym}$ is the symmetrization operator $N!^{-1}\sum_{P}P$ over all the permutation of $N$ particles. The operator  $\hat{U}_{\vec\theta}$ can be a linear gate such as phase-shifter or beam splitter, or non-linear such as a Kerr gate, a non-linear phase-shifter or a more complex photon-preserving transformation.  At that point, $\vert \psi_i \rangle$ can be either a single particle state or a multi-particles state.
In \efig{fig_circuit}, the arrows correspond to quantum modes of light $\vert 0\rangle$, $\vert 1\rangle$,...,$\vert d\rangle$ that enter the system and $\mathbb{P}_{\vec\theta}$ is the result of a photon measurement protocol which will be defined later on.
For the sake of the discussion, we will only consider the case of linear phase-shifters, the transformation acting on quantum mode $i$ is given by:
\[
\hat{U}_\theta = e^{i\theta \hat{a}_i^\dagger \hat{a}_i}.
\]
Here, $\hat{a}_i^{\dagger}$ and $\hat{a}_i$  are the creation and annihilation operators for mode $i$ and $\theta$ is the value of phase shift. 
The operator $\hat{U}_0$ can be any generic unitary operator and the operator $\hat{U}_{\vec\theta}$  describes multiple phase-shifters on different modes $\vec\theta=[\theta_1,\theta_2...\theta_{n_{\mathrm{PS}}}]$. In the following, let us model a generic random unitary evolution by a Haar operator
\begin{equation}\label{haar}
\hat{U}_{\mathrm{Haar}}\in \mathrm{U}(d).
\end{equation}
The Haar random unitary operator is interesting because it is uniformly distributed over the space of all unitary matrices in dimension $d$. This random operator is not only the most general form of quantum unitary evolution, it is also experimentally realizable with simple optical components such as beam-splitter and phase-shifters \cite{de2018simple}. The interest in quantum random dynamics has emerged recently as a fantastic playground for quantum many-body physics, especially in the hope to better understand universal properties of entanglement and thermalization in complex quantum systems (see \cite{potter2022entanglement,fisher2023random,skinner2023lecture} for recent reviews)
 
We choose here to work with several types of entangled initial states $\vert \psi_i \rangle$, that will be defined in this section. Let us first use the very general notion of Schmidt rank as a measure of entanglement \cite{sperling2011schmidt} to quantify the complexity of the initial states. The bipartite quantum state $\vert \psi_i \rangle$ is said to be entangled if its Schmidt rank (i.e. number of singular values of the Schmidt decomposition) is strictly greater than $1$, and is not entangled otherwise. As a reminder, for any quantum state $\vert \psi_i \rangle$ acting on subsystems $A$ and $B$ there is some set of orthonormal states $\vert A\rangle$ and $\vert B\rangle$ such that $\vert \psi_i \rangle$ can be written as
\begin{equation}\label{schmidt}
\vert \psi_i \rangle=\sum_{k=1}^{d} \Lambda_k \vert A\rangle\otimes\vert B\rangle
\end{equation}
with $\sum_{k}\Lambda_k^{2}=1$. The Schmidt rank is the number of non zero element in that sum, for separable state $\mathrm{rank}(\vert \psi_i \rangle)=1$ and the maximal rank is  $\mathrm{rank}(\vert \psi_i \rangle)=d$. The coefficients are related to the eigenvalues $\lambda_k$ of the reduced density matrix $\rho_A$ by $\lambda_k=\Lambda^{2}_k$ such that $\sum_{k}\lambda_k=1$. The $n$-th Renyi entropy can be written as
\begin{equation}
S_n=\frac{1}{1-n}\mathrm{log}\sum_{k=1}^{d}\lambda_k^{n}.
\end{equation}
One implication is that the zeroth Renyi entropy $S_0$ is exactly given by the logarithm of the number of terms in the Schmidt decomposition
\begin{equation}
S_0=\mathrm{log}(\mathrm{rank}(\vert \psi_i \rangle)).
\end{equation}
That is, $S_0$ counts the minimal number of terms that you need in order to write a decomposition of the form of \eref{schmidt} if you want to represent the state exactly.
To illustrate the problem, let us examine in the rest of the paper two initial states that have very different entanglement structure. The first one is a weakly entangled initial state defined by a superposition of coherent states
\begin{equation}\label{weak}
\vert \psi_i \rangle=|c_{\alpha_1},c_{\alpha_2}\rangle_{\mathrm{sym}}  \ \ \ \ \mathrm{rank}(\vert \psi_i \rangle)=2,
\end{equation}
where $|a,b\rangle_{\mathrm{sym}}:=\sqrt{C}\big(|a,b\rangle+|b,a\rangle\big)$ with $\sqrt{C}$ is a normalization constant. Let's choose $\alpha_1=2,\alpha_2=3$ in the rest of the paper.  The (non-normalized) truncated coherent state $|c_\alpha\rangle$ in dimension $d$ is characterized by a complex parameter $\alpha$ and can be written as:
\begin{equation}\label{cs}
|c_{\alpha}\rangle = e^{-\frac{|\alpha|^2}{2}} \sum_{m=0}^{d} \frac{\alpha^m}{\sqrt{m!}} |m\rangle,
\end{equation}
where $|m\rangle$ represents photonic modes (spatial modes of photons) with $d$-states \cite{fabre2020modes}, and $\alpha$ is a complex number that determines the properties of the coherent state. This state is expressed in the basis of the quantum modes $m$ not occupation numbers as it is usually defined.
The choice of this specific initial state is rather arbitrary and only satisfies the minimal rank condition. At the opposite end of the spectrum, the maximally entangled 2-photon state takes the form
 \begin{equation}
\vert \psi_i \rangle=\frac{1}{\sqrt{d}}\sum_{m=1}^{d} |m,m\rangle , \ \ \ \ \mathrm{rank}(\vert \psi_i \rangle)=d.
\end{equation}
Despite its less than usual looking form, this state is the generalized $\mathrm{NOON}$ state (with $N=2$ and $d$ quantum modes), often used to show quantum advantage in the field of quantum imaging and sensing. We can already notice that the entropy $S_0$ is constant for any $d$ for the weakly entangled state and scales with $\mathrm{log}(d)$ for the NOON state. The choice of weakly/strongly entangled states will become clear in the following when performing quantum parameter estimation. With that in mind, let us define $\vert \psi_{\vec\theta}\rangle = \mathrm{Sym}(\hat{U}^{\otimes 2}\vert \psi_i \rangle)$ with $\hat{U}=\hat{U}_{0}.\hat{U}_{\vec\theta}$ and $\vec\theta=[\theta_1,\theta_2...\theta_6]$ and all other phase-shifters set to zero.
 Finally the measurement process showed in \efig{fig_circuit} is a full coincidence measurement defined by 
\begin{equation}
\big[\mathbb{P}_{\vec\theta}\big]_{ij}= |\langle\psi_{\vec\theta} | m_i,m_j\rangle_{\mathrm{sym}}|^{2},
\end{equation}
where $\big[\mathbb{P}_{\vec\theta}\big]_{ij}$ is the matrix element equal to the coincidence probability that one photon is detected in mode $m_i$ and the other in mode $m_j$ such that $\sum_{ij}\big[\mathbb{P}_{\vec\theta}\big]_{ij}=1$. After reshaping the probability vector into a matrix of shape $d$ by $d$, the probability of coincidence for completely indistinguishable bosons can be defined on the lower hyper-triangle by a triangularization procedure. This idealized setup is not only a theoretical toy-model, since an experimental 2-photon (with $d=16$) time-resolving coincidence detector has successfully been developed using superconducting nanowires \cite{zhu2018scalable}. 

\begin{figure}[ht!]
\includegraphics[scale=0.13]{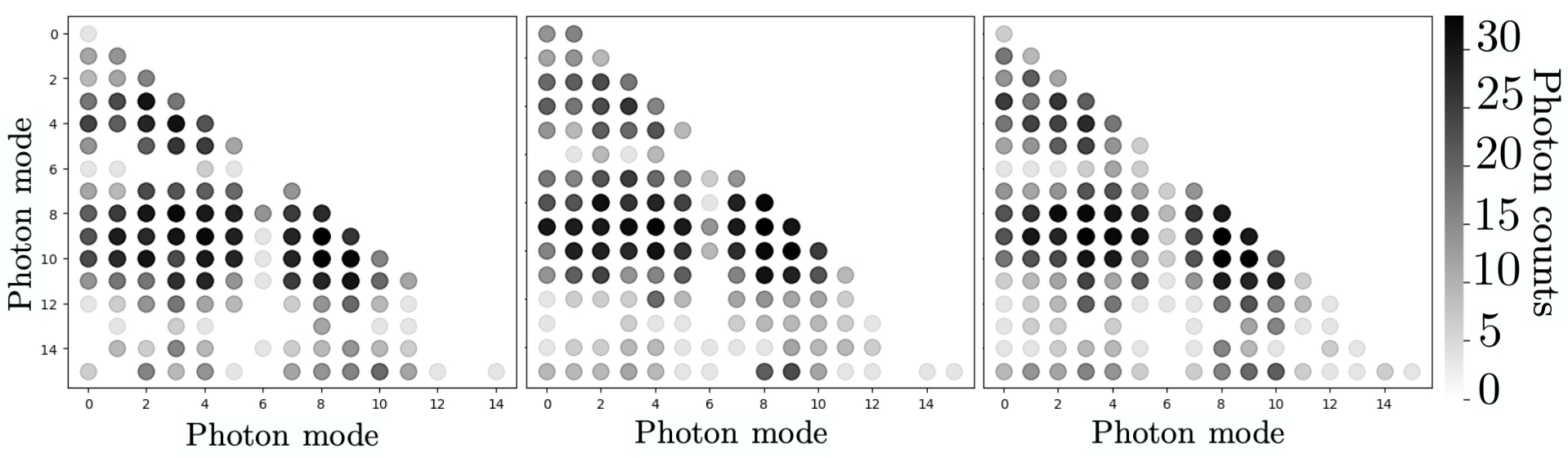} 
	\caption{Examples of photon coincidence detection patterns with $\vert \psi_i \rangle=|c_{2},c_{3}\rangle_{\mathrm{sym}}$, $d=16$, $p=1000$ and the same Haar matrix and $\vec\theta$.}
	\label{fig_coincidence_coherent}
\end{figure}

\begin{figure}[ht!]
\includegraphics[scale=0.11]{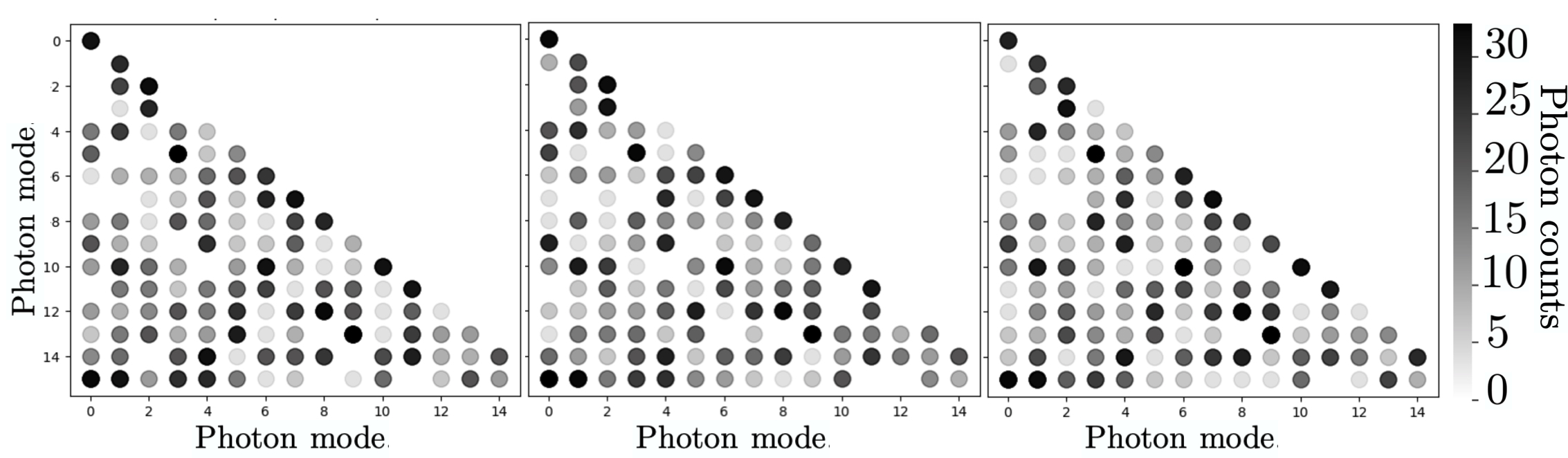} 
	\caption{Examples of photon coincidence detection patterns with $\vert \psi_i \rangle=d^{-1/2}\sum_{m=1}^{d} |m,m\rangle$, $d=16$, $p=1000$ and the same Haar matrix and $\vec\theta$.}
	\label{fig_coincidence_NOON}
\end{figure}

Let us sample this probability $p$ times, corresponding to repeating the experiment multiple times and measuring the coincidence patterns. Examples of detection patterns resulting from the two states are showed on \efig{fig_coincidence_coherent} and \efig{fig_coincidence_NOON}. The resulting normalized probability distribution is noted $\mathbb{P}_\mathrm{emp}(p)$ and in our simulations, the only noise is due to the photon sampling parameter $p$, such that at the limit $p\rightarrow\infty$ the empirical probability converges to the exact probability $\mathbb{P}_\mathrm{emp}(p)\rightarrow \mathbb{P}_{\vec\theta}$. 
Looking at the figures above, we can clearly see a difference in structure between the two final configurations for the same unitary operator $\hat{U}_{\mathrm{Haar}}$. While the weakly entangled state (\efig{fig_coincidence_coherent}) produces fairly structured patterns, the strongly entangled state (\efig{fig_coincidence_NOON}) creates a lot more randomness in the probability distributions. The qualitative understanding of the entanglement growth will show its importance later on in the paper, when the low-rank structure of the states will be taken into account for better classical simulations.

\section{Quantum estimation and back-propagation}

One of the most studied metrology problem that comes up in numerous applications of quantum optics is the following \cite{humphreys2013quantum,pezze2014quantum,liberman2016quantum,szczykulska2016multi}: can we estimate (without any prior distribution) the values of the angles $\vec{\theta}$  of phase-shifters (\efig{fig_circuit}) given a coincidence pattern, the unitary operator and the initial state. The problem is summarized as
\begin{equation}
\boxed{\mathbb{P}_\mathrm{emp}(p)\rightarrow \vec\theta \ \ \mathrm{for} \ \ p\ll\infty.}
\end{equation}
The question is highly ambiguous, as there are potentially many solutions for a given probability measurement. The design of an initial state and measurement protocols that make the problem uniquely defined and optimal is a very complicated task and part of the field of quantum tomography. We let that complication aside as it is not the purpose of this discussion and only choose cases without too much ambiguities.

Popular methods of quantum parameter estimation include (among others) bayesian and gradient-based methods. The quantum evolution \eref{heisenberg}, mathematically described through basic linear algebra in Hilbert space, is very well suited to gradient descent via back-propagation. Back-propagation is a technique widely employed in deep learning learning frameworks (such as $\mathrm{Tensorflow}^\copyright$ , $\mathrm{PyTorch}^\copyright$ or $\mathrm{Jax}^\copyright$) to efficiently compute gradients of a computational graph. It allows to automatically and exactly calculate derivatives of a function with respect to its input parameters, which is the corner stone of gradient-based optimization algorithms used in modern deep learning. We do not detail the ins and outs of these methods and send the reader to the standard literature on machine learning (ex \cite{goodfellow2016deep}).

The loss function being used for the optimization of the unitary evolution \efig{fig_circuit} is the Kullback-Leibler ($\mathrm{KL}$) divergence defined by
\begin{equation}\label{KL}
\mathrm{KL}\big(\mathbb{P}_{\tilde{\vec\theta}} |\mathbb{P}_\mathrm{emp}(p)\big)=-\Big\lVert \mathbb{P}_{\tilde{\vec\theta}} \ \mathrm{log}\Big(\frac{\mathbb{P}_\mathrm{emp}(p)}{\mathbb{P}_{\tilde{\vec\theta}} }\Big)\Big\rVert,
\end{equation}
where $\lVert.\rVert$ is the matrix sum and $\tilde{\vec\theta}$ is the estimated phase-shift vector. A gradient descent algorithm that minimizes the $\mathrm{KL}$ can be implemented. We choose the Adam algorithm \eref{adam} which maintains two moving averages for each parameter: the first moment estimate and the second moment estimate. These moving averages are denoted as \(m_t\) and \(v_t\) respectively.
\begin{figure}[ht!]
\includegraphics[scale=0.22,left]{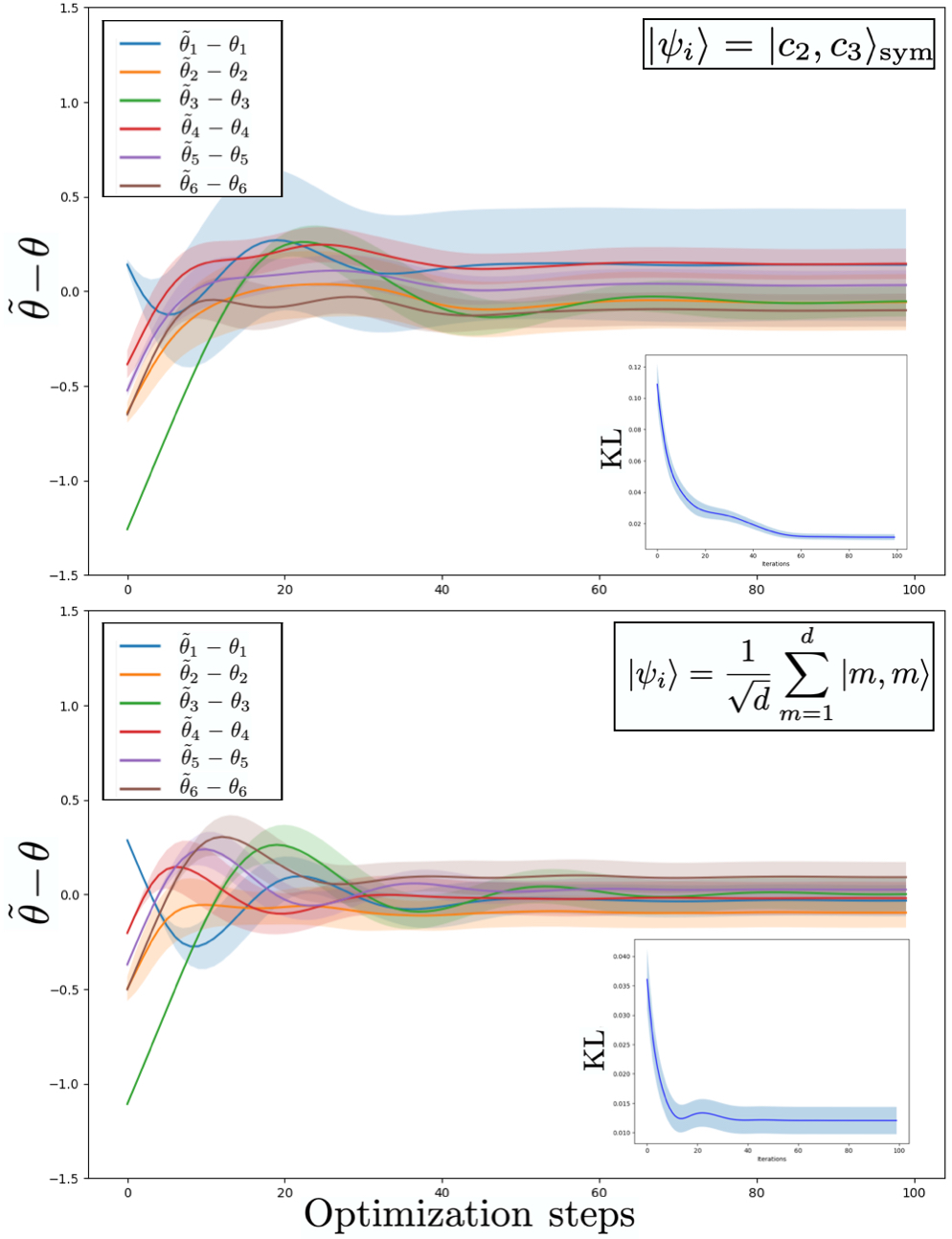} 
	\caption{Results of optimizations of the exact unitary evolution in $d=16$ with 100 different $\mathbb{P}_{\mathrm{emp}}(p=1000)$ with $\vert \psi_i \rangle=|c_{2},c_{3}\rangle_{\mathrm{sym}}$ (Top) and the $\mathrm{NOON}$ state (bottom): the figures show the convergence of the mean phase shift residuals and their standard deviations over the 100 trials. The difference in precision between the two initial states is evident and explained in the main text.}
\label{fig_opt}
\end{figure}
The algorithm updates the parameters $\tilde{\theta}$ at each iteration using the following formula
\begin{equation}\label{adam}
\tilde{\theta}(k+1) = \tilde{\theta}(k) - \alpha\frac{\hat{m}(k)}{\sqrt{\hat{v}(k)} + \epsilon},
\end{equation}
with $\hat{m}(k)=\frac{m(k)}{1 - \beta_1^k}$ and $\hat{v}(k)=\frac{v(k)}{1 - \beta_2^k}$ and
\[
\begin{aligned}
m(k)&= \beta_1 \cdot m(k-1) + (1 - \beta_1) \cdot \nabla \mathrm{KL}_{\tilde{\theta}}, \\
v(k) &= \beta_2 \cdot v(k-1) + (1 - \beta_2) \cdot (\nabla \mathrm{KL}_{\tilde{\theta}})^2.
\end{aligned}
\]
Here, $\tilde{\theta}(k)$ is the parameter at step $k$, $\nabla \mathrm{KL}_{\theta}$ is the gradient of the loss function with respect to the phase shifts at step \(k\), \(\alpha\) is the learning rate. \(\beta_1\) and \(\beta_2\) are smoothing parameters,
\(m(k)\) and \(v(k)\) are the first and second moment estimates.
\(\hat{m}(k)\) and \(\hat{v}(k)\) are the bias-corrected estimates, and
\(\epsilon\) is a small constant to avoid division by zero. The back-propagation estimation can be summarized as follow
\begin{equation}
\boxed{\mathrm{Find} \ \tilde{\vec\theta} \ \mathrm{such \ that} \ \mathrm{KL}\big(\mathbb{P}_{\tilde{\vec\theta}} |\mathbb{P}_\mathrm{emp}(p)\big) \ \mathrm{is \ min}.}
\end{equation}
Results of $100$ optimizations with different $\mathbb{P}_{\mathrm{emp}}(p=1000)$ are presented on \efig{fig_opt}, the estimated $\tilde{\theta}_i$ 'converge' towards values of the phase shifts that minimize the loss function \eref{KL}. The results of \efig{fig_opt} confirm a very well known feature of quantum metrology, that the strength of entanglement helps make for better quantum estimation.
 It has been showed again and again \cite{nagata2007beating,slussarenko2017unconditional,demkowicz2012elusive}, that NOON states and relatives are ultimate probes for quantum metrology and sensing, due to their ability to reach the Heisenberg limit of precision. The predictive capability of the system depends greatly on the structure of entanglement and can (in some cases) be computed exactly using the quantum Fisher information (see \cite{barbieri2022optical} for details). 
While we have performed the task of quantum parameter estimation using back-propagation, computation quickly becomes intractable when $d$ increases. Furthermore, whereas our paper concentrates on a quantum system readily adaptable to back-propagation, not all systems share this characteristic, making such methods inadequate in those circumstances. In the following we will show how to replace the exact unitary evolution by a parameterized network evolution as a proxy in order to perform the back-propagation optimization in a smaller space, allowing us to access higher $d$ at the price of a small precision cost and time-consuming supervised training. Also we will see how the architecture of the network can be adapted to the entanglement structure of the final states and by doing so, can help simulate quantum systems efficiently.

\section{Quantum consistent neural network for weakly/strongly entangled states}

\begin{figure}[ht!]
\includegraphics[scale=0.135,left]{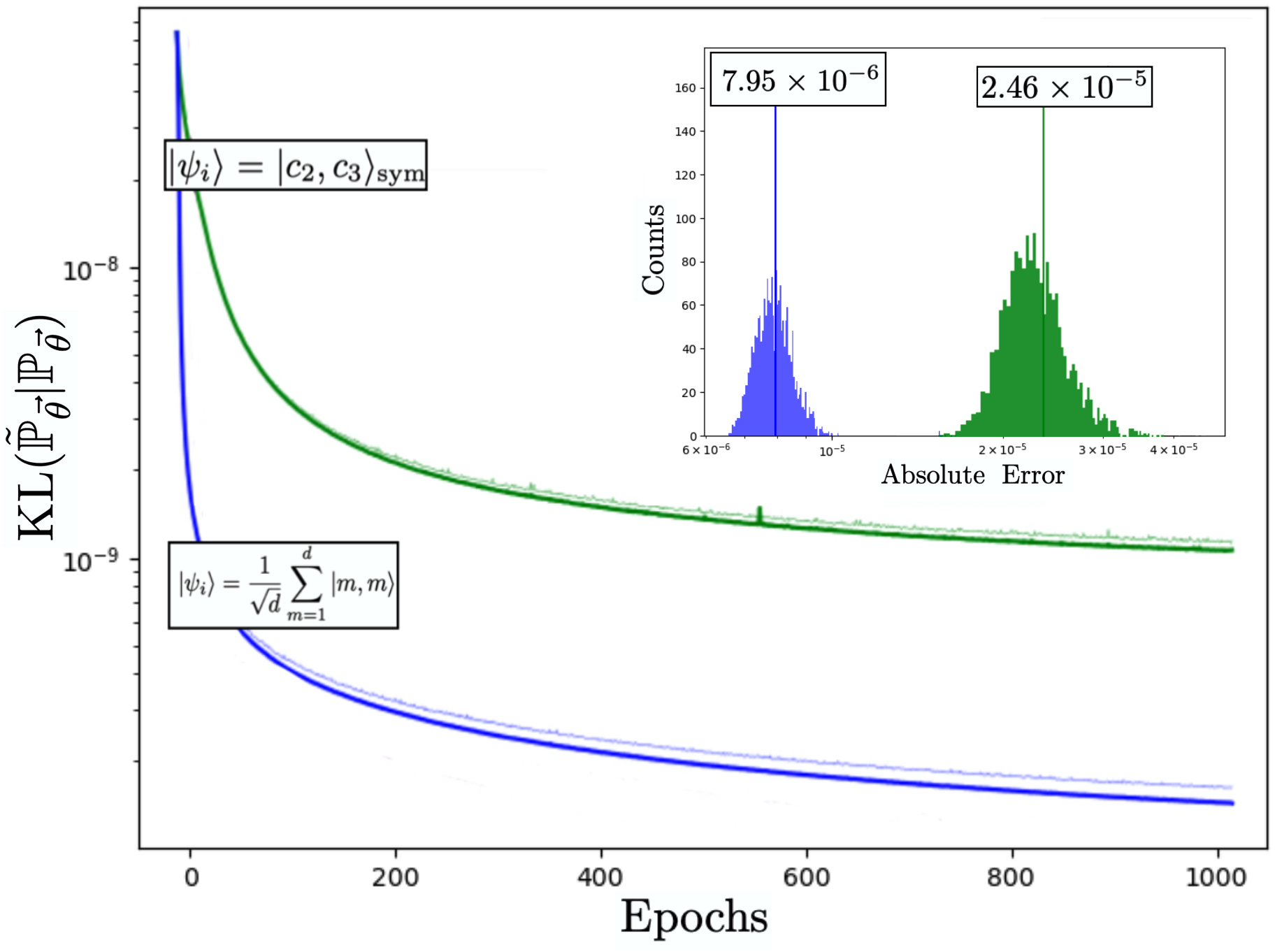} 
	\caption{\textcolor{black}{Strongly (blue) vs weakly (green) entangled trainings with the quantum consistent neural net. The plain color are the training losses whereas the shadowed lines are the corresponding validation losses. The absolute error histograms are plotted as well as their mean.}}
\label{fig_training_0}
\end{figure}

As mentioned in the introduction, the fast moving field of machine learning is deeply and gradually influencing research in quantum physics, allowing new methods to tackle hard problems and helping shape new technologies, especially in the field of quantum computing, through better data analysis and simulations.  
In the following we propose to explore this route and use supervised deep learning to model the unitary dynamics \eref{heisenberg}. 
The first step is to generate a proper dataset, \textit{ie.} by generating $N_{\mathrm{label}}=10000$ examples of $\mathbb{P}_{\vec\theta}$ with phase-shifts vectors $\vec\theta$ with ($n_{\mathrm{PS}}=6)$ chosen randomly between $[0,2\pi]^{\times n_{\mathrm{PS}}}$ and $d=64$. As it is customary in machine learning, the dataset is divided into training and validation sets (with a percentage of 70/30) such that the learning is done using solely the training set.  The task at hand is to replaced the full unitary evolution by a neural net that estimates $\mathbb{P}_{\vec\theta}$ such that 
 \begin{equation}
\boxed{\tilde{\mathbb{P}}_{\vec\theta}=\mathrm{NN}_{\vec w}(\vec\theta),}
\end{equation}
where  $\mathrm{NN}_{\vec w}$ is a neural network parameterized by the weights $\vec w$ and optimized by supervised learning using the previously defined dataset. Let us note that we could have used the neural net to model directly the final state (such that $|\psi_\theta\rangle\approx\mathrm{NN}_{\vec w}(\vec\theta)$) and not the probabilities, as it is generally done in the field of quantum state learning, but we want this method to be applicable in experimental settings where only measurements are available. The loss function for the network training is similar to the one used previously for optimization, precisely
\begin{equation}\label{KL_loss}
\mathcal{L}_{\vec w}=\sum_{i\in\{\mathrm{batch}_p\}}\mathrm{KL}\big(\mathrm{NN}_{\vec w}(\vec\theta) |\mathbb{P}_\mathrm{emp}(p)\big),
\end{equation}
where $\mathrm{batch}_p$ is a random set of $p$ elements of the training set and details about the supervised training will be explained later on. There is considerable freedom in designing a suitable neural network $\mathrm{NN}_{\vec w}(\vec\theta)$; however, our objective here is to craft a network (quantum consistent neural net \footnote{This naming can be a little confusing as this network is not what is colloquy called a quantum neural network, but it is nonetheless a neural network that happens to be quantum consistent.}) that aligns as closely as possible with the quantum system. Incorporating physics into the architecture and/or the training of neural networks is becoming more and more popular in the scientific machine learning community, either by enforcing symmetries and invariances \cite{greydanus2019hamiltonian,cranmer2020lagrangian} or directly by using the ordinary or partial differential equations if available \cite{karniadakis2021physics,cuomo2022scientific}. It can be argued that the inclusion of physical laws becomes even more crucial for quantum systems in which every bit of prior information can be useful to mitigate the exploding growth of simulation overhead. In our present case, the quantum consistency of the machine learning architecture ensures that parameter estimation operates in the neural net space similarly as it would in the full Hilbert space. For this to be achieved, we first need to encode the shift vector $\vec\theta$ in a circular invariant space.
\begin{figure}[ht!]
\includegraphics[scale=0.165]{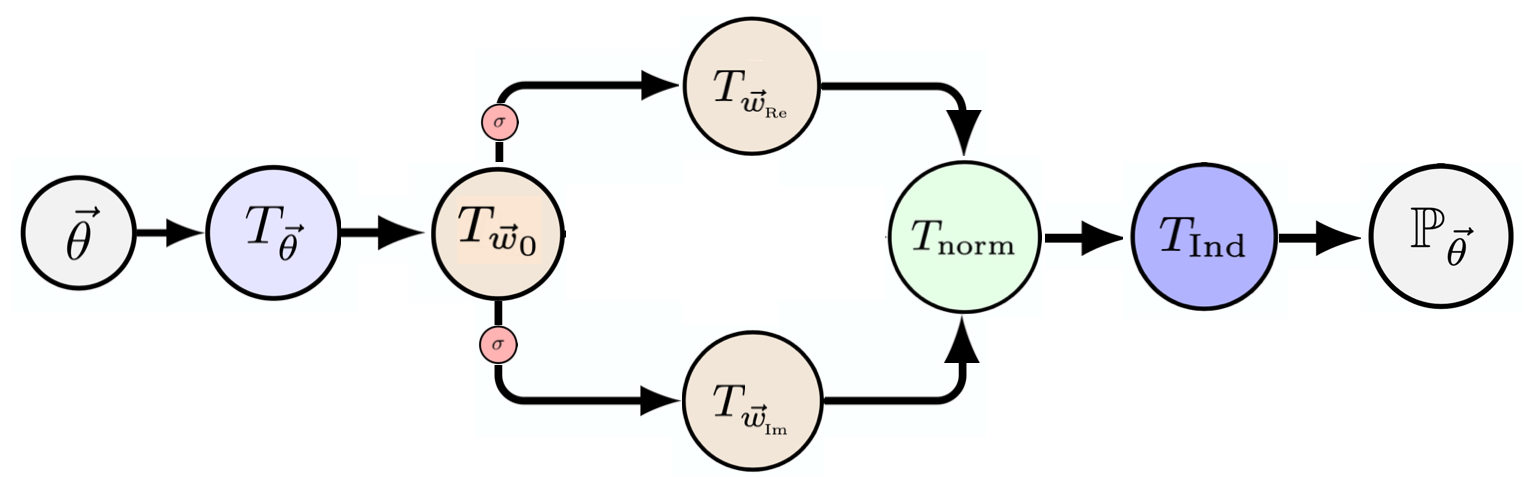} 
	\caption{Representation of the layers of the quantum consistent neural network. Starting from the vector $\vec{\theta}$ in input, the system performs a sequence of transformations and outputs a coincidence probability matrix $\mathbb{P}_{\vec\theta}$. The successive layers are described in the main text.}
\label{fig_network}
\end{figure}
The invariance of the shift operators $\hat{U}_{\theta}= \hat{U}_{\theta+2p\pi}$, can be encoded by using the following feature map
\begin{equation}\label{U1}
T_{\vec\theta}(\vec\theta)=(\mathrm{cos}(\vec\theta),\mathrm{sin}(\vec\theta))^{T}=\overset{\curvearrowleft}{\theta},
\end{equation}
such that $T_{\vec\theta}=T_{\vec\theta+2p\pi}$. The transformations $T_{\vec{w}_0}$ and $T_{\vec{w}_\mathrm{Re/Im}}$ in \efig{fig_network} represent standard feed-forward layers. In the first one, the output of the parameterized layer $T_{\vec{w_0}}$ is obtained by applying a linear transformation followed by a non-linear activation function $\sigma$
\begin{eqnarray}\label{perceptron}
&&T_{\vec{w_0}}(\overset{\curvearrowleft}{\theta})=\mathbf{W}_{0}\overset{\curvearrowleft}{\theta},\nonumber\\
&&\sigma\big(\mathbf{W}_{0}\overset{\curvearrowleft}{\theta}\big):=\vert\chi_\theta \rangle,
\end{eqnarray}
where $\vert\chi_\theta \rangle$ is a circularly invariant real valued vector of length $l$ (we choose $l=100$ in the following). Here, $\mathbf{W}_{0}$ represents the map from a vector of size $2n_{\mathrm{PS}}$ into one of size $d\times d$ with $n_{\mathrm{PS}}=6$. The number of parameter of this transformation is $\#(T_{\vec{w_0}})=2n_{\mathrm{PS}} l$.  The activation function (rectified linear function $\sigma(x)=\mathrm{max}(0,x)$) is applied to each element of the resulting vector.  In the subsequent parameterized layers, we choose to separately and independently model the real and imaginary parts of the state with $T_{\vec{w}_\mathrm{Re}}$ and $T_{\vec{w}_\mathrm{Im}}$ such that
\begin{eqnarray}\label{perceptron_psi}
T_{\vec{w}_\mathrm{Re}}\vert\chi_\theta \rangle&=&\mathrm{Resh}_{(d,d)}\mathbf{W}_{\mathrm{Re}}\vert\chi_\theta \rangle + (\mathrm{Resh}_{(d,d)}\mathbf{W}_{\mathrm{Re}}\vert\chi_\theta \rangle)^{T}\nonumber\\
&:=&\mathrm{Re}\vert\psi_\theta\rangle\nonumber,\\
T_{\vec{w}_\mathrm{Im}}\vert\chi_\theta \rangle&=&\mathrm{Resh}_{(d,d)}\mathbf{W}_{\mathrm{Im}}\vert\chi_\theta \rangle + (\mathrm{Resh}_{(d,d)}\mathbf{W}_{\mathrm{Im}}\vert\chi_\theta \rangle)^{T}\nonumber\\
&:=&\mathrm{Im}\vert\psi_\theta \rangle,
\end{eqnarray}
with $\mathrm{Resh}_{(d,d)}$ reshapes vectors of length $d^2$ into $d$ by $d$ matrices. At that point, the symmetry of the real and imaginary part of the wave function is enforced and $\mathbf{W}_{\mathrm{Re/Im}}$ represent maps from vectors of length $l$ into lenght $d^2$, such that the number of parameters is $\#(T_{\vec{w}_\mathrm{Re}})=\#(T_{\vec{w}_\mathrm{Im}})=l\times d^2$.
 The resulting wave-function is a matrix instead of a vector, all the subsequent transformation will act on matrices. Let us note that there is no need for activation function at the output of this layer, since the following normalization layer will act as a nonlinear activation.  The normalization of the probability can be hard-wired inside the network by using a Boltzmann normalization $T_{\mathrm{norm}}(X,Y)$  that takes in entry the two pathway outputs and preserves probability of the wave-function
\begin{eqnarray}\label{boltzmann}
&&T_{\mathrm{norm}}\big(\mathrm{Re}\vert\psi_\theta\rangle,\mathrm{Im}\vert\psi_\theta\rangle\big)\nonumber\\
&&=Z^{-1}\exp\Big(-\beta\Big\lVert |\mathrm{Re}\vert\psi_\theta\rangle+i\mathrm{Im}\vert\psi_\theta\rangle|^{2}\Big\lVert\Big),
\end{eqnarray}
with the denominator being a normalization constant, such that $\Big\lVert T_{\mathrm{norm}}\big(x,y\big)\Big\lVert=1$ for all $x,y$. The inverse temperature $\beta$ controls the level of stochasticity in the choice, ranging from $\beta=0$ for completely uniform distribution and $\beta=\infty$ for deterministically choosing the highest value. The value $\beta=1$ is often use in machine learning (it is then called the softmax activation function), but here a larger value of $\beta=1000$ can be set as the system is highly constrained by the physical structure, and parameter exploration is not nearly as needed as rapid cooling down. Finally, triangularizing the coincidence probability matrix guarantees photon indistinguishability
\begin{eqnarray}\label{triangle}
T_{\mathrm{ind}}\big(T_{\mathrm{norm}}(\mathrm{Re}\vert\psi_\theta ),\mathrm{Im}\vert\psi_\theta ))\big)=\mathbb{P}_{\vec\theta},
\end{eqnarray}
with
\begin{eqnarray}
T_{\mathrm{ind}}(X)=X.X^{T}.L_{d} (\mathbb{I}_d-\mathrm{diag}_{d})+X.\mathrm{diag}_{d},
\end{eqnarray}
where $L_{d}$ is the lower triangular unity matrix and $\mathrm{diag}_{d}$ is the diagonal matrix in dimension $d$. 
In a nutshell, the output of $T_{\mathrm{ind}}$ should be invariant by $2p\pi$, lower triangular and of unit norm. The three matrices ($\mathbf{W}_0$, $\mathbf{W}_{\mathrm{Re}}$ and $\mathbf{W}_{\mathrm{Im}}$) contain all the free parameters of the neural network and will be optimized by gradient descent. The physical layers ($T_{\vec{\theta}}$, $T_{\mathrm{norm}}$ and $T_{\mathrm{ind}}$) are parameter-free and crucial to enforce a physically consistent result needed for quantum parameter optimization.
 The final quantum consistent neural net is showed on \efig{fig_network}. Training is done using the ADAM algorithm \eref{adam} with a batched generalization (batch size of 32) of the $\mathrm{KL}$ divergence \eref{KL}, and learning rate set to $\alpha=0.1$. 
 \begin{table}[ht!]
\centering
\begin{tabular}{ ||c | c|c| c || } 
\hline
 & $\mathrm{MAE(Strong)}$ & $\mathrm{MAE(Weak)}$ \\
\hline
$\mathrm{QCNN}$ & $7.95\times 10^{-6}$ &$2.46\times 10^{-5}$  \\
Vanilla NN & $2.50\times 10^{-5}$ &$7.95\times 10^{-5}$ \\
\hline
\end{tabular}
\caption{Mean absolute errors of the $\mathrm{QCNN}$ vs a vanilla neural net for both initial states. The improvement is approximatively consistent among both cases.}
\label{table}
\end{table}
No other machine learning tricks such as dropouts or regularization are used and the training and the histograms of errors are plotted in blue on \efig{fig_training_0}. Comparisons of mean absolute errors with a simple single layer neural net with no physical layers are showed on the table above for both the strongly and weakly entangled states. A rather consequent improvement is achieved by the $\mathrm{QCNN}$ on both states due to the physical structure of the architecture \efig{fig_network}.

Despite showing both very good results, we observe in \efig{fig_training_0}, that the $\mathrm{QCNN}$ approximates significantly better strongly entangled states that weaker ones. The difference in mean absolute error is 4 times smaller for the NOON state compared to the weakly entangled coherent state. This was to be expected, as it is well known since the seminal paper of Carleo and Troyer \cite{carleo2017solving} that neural quantum states (NQS) are excellent candidates for complex states that do not obey area laws, such as strongly entangled states or critical points (also see \cite{lange2024architectures} for a recent review). With that being said, we will see how to modify the parameterized operations of the network to take into account the structure of low entanglement/low rank of the initial states, by replacing matrix-vector multiplications such as \eref{perceptron} and \eref{perceptron_psi} by tensor-matrix contractions. We demonstrate how tensor networks and matrix product states, objects that are known to describe (sometimes exactly) correlated quantum and statistical physics systems very efficiently and accurately, can outperform significantly neural networks in case of low-rank quantum states and mitigate the computation burden of learning high-dimensional quantum systems.

\section{Low rank states, tensor networks and parameter estimation}
\textcolor{black}{Much like neural networks serve as universal approximations for general distributions, tensor networks, in principle, can model any discrete distribution given sufficiently big bond dimensions \cite{stoudenmire2016supervised,glasser2018supervised}, in particular strongly correlated quantum systems in which the Hilbert space structure enables the utilization of global internal symmetries and specific entanglement properties. This becomes crucial when efficiently expressing probability distributions with invariances and correlations resulting from such properties. In this section, we will show how to incorporate such structures into the general architecture \efig{fig_network} and see how it compares to the neural net introduiced previously.} 

\begin{figure}[ht!]
\includegraphics[scale=0.14]{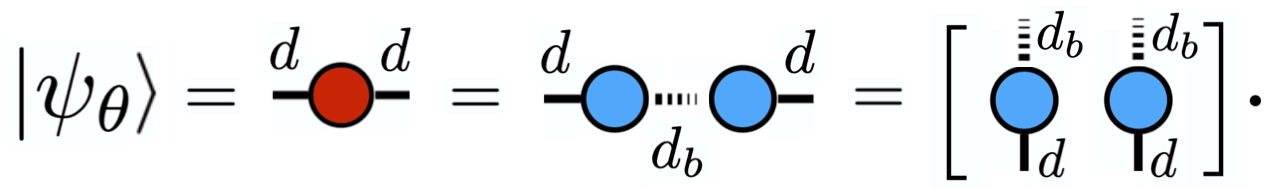} 
\caption{Matrix product decomposition of the wave-function in Penrose notation: The quantum state $|\psi_\theta)$ of dimension $(d\times d)$ can be written as a product of smaller two $(d\times d_{b})$ matrices. As a reminder, circles with $p$ lines represent tensors of dimension $p$, contraction of tensors are represented by joined lines, and separated circles represent the outer product. Shapes, colors and geometry are irrelevant except stated otherwise.}
\label{fig_mps_0}
\end{figure}

\begin{figure}[ht!]
\includegraphics[scale=0.15]{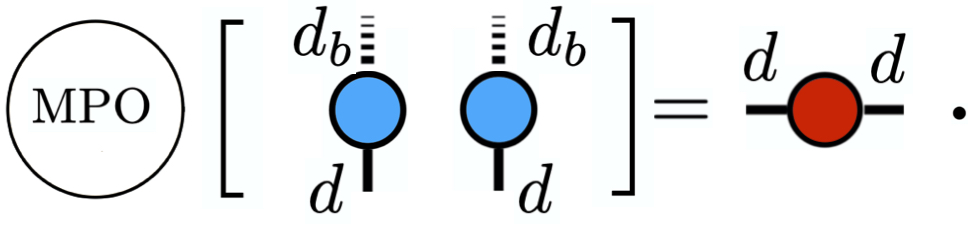} 
\caption{Matrix product operator that contains all the parameterized weights. The MPO will then be contracted with the MPS to recover the full $d\times d$ matrix.}
\label{fig_mpo}
\end{figure}

\begin{figure}[ht!]
\includegraphics[scale=0.18]{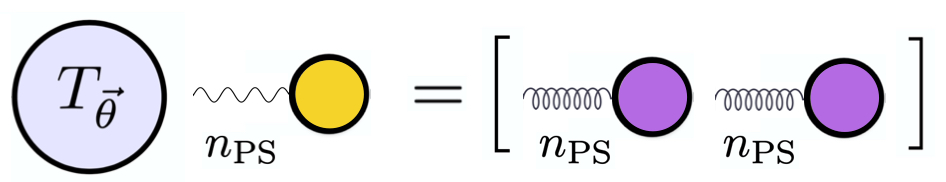} 
\caption{Periodic feature space. The springy links represents the phase dimension $n_{\mathrm{PS}}$ and the two vector correspond to $\cos(\vec\theta)$ and $\sin(\vec\theta)$.}
\label{fig_contraction_theta}
\end{figure}

\begin{figure}[ht!]
\includegraphics[scale=0.16]{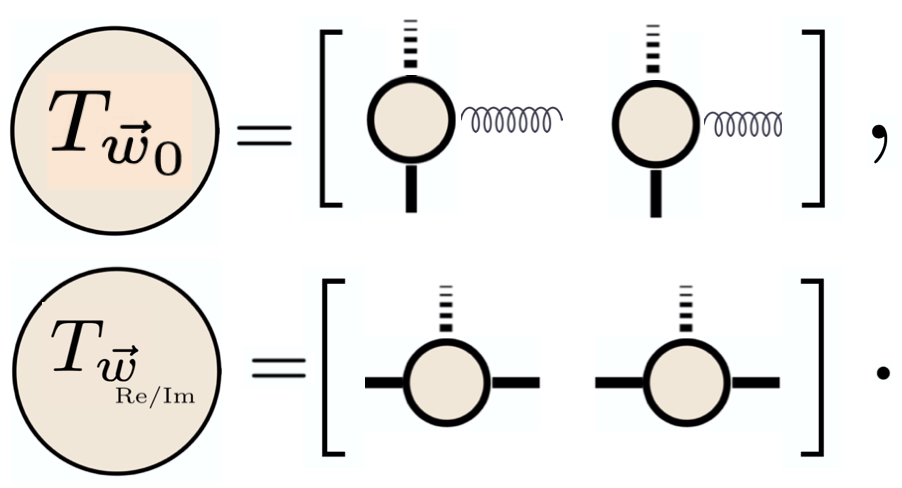} 
\caption{Parameterized matrix product operators. The dotted lines represent the bond dimension $d_{b}$ of the MPO and match the bond dimensions of the MPS.}
\label{fig_contraction_0}
\end{figure}

\begin{figure}[ht!]
\includegraphics[scale=0.16]{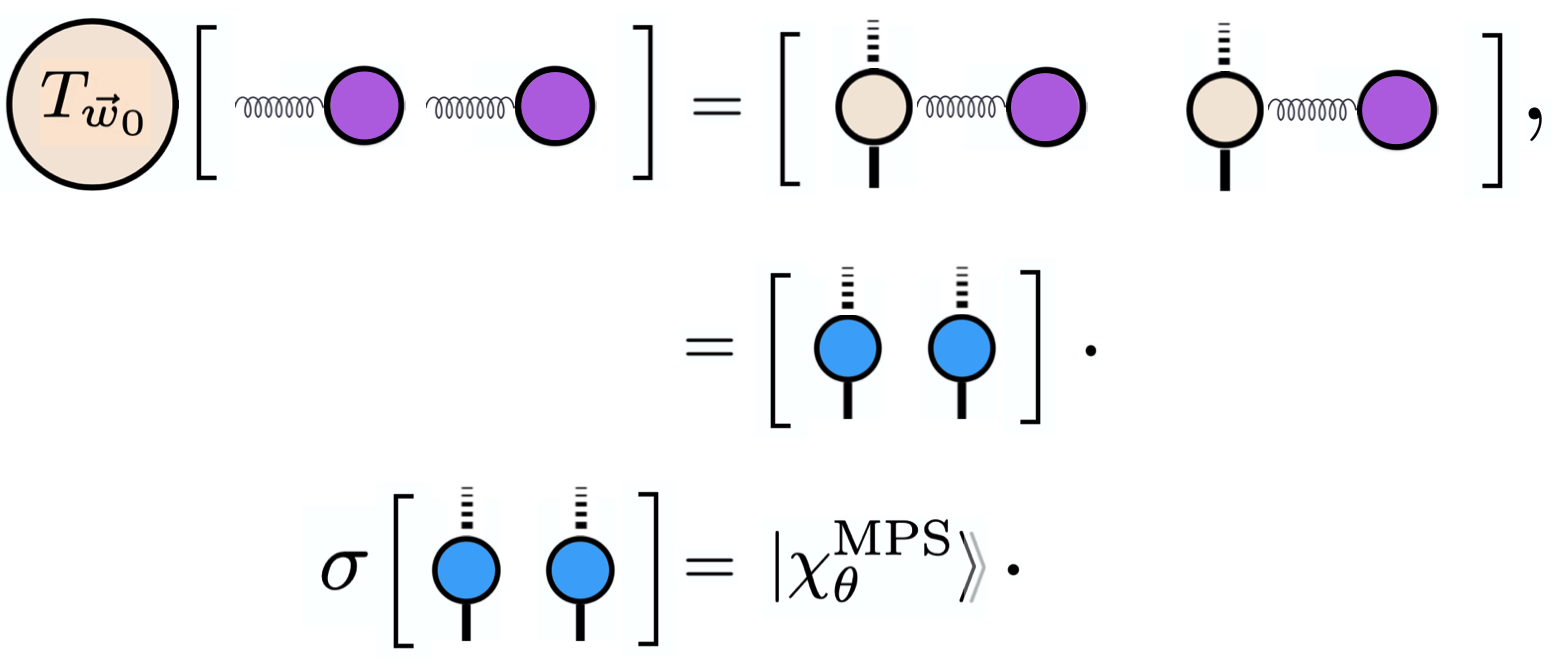} 
\caption{The feed-forward transformation $\mathbf{W}_0$ used previously in \eref{perceptron} is replaced by a simple tensor-vector contraction that generates a matrix product state $\vert\chi^{\mathrm{MPS}}_\theta \rangle\rangle$. The dashed lines are of dimension $d_{b}$, the plain lines of dimension $d$ and the springy lines of dimension $n_{\mathrm{PS}}$.}
\label{phase_contraction_1}
\end{figure}

\begin{figure}[ht!]
\includegraphics[scale=0.14]{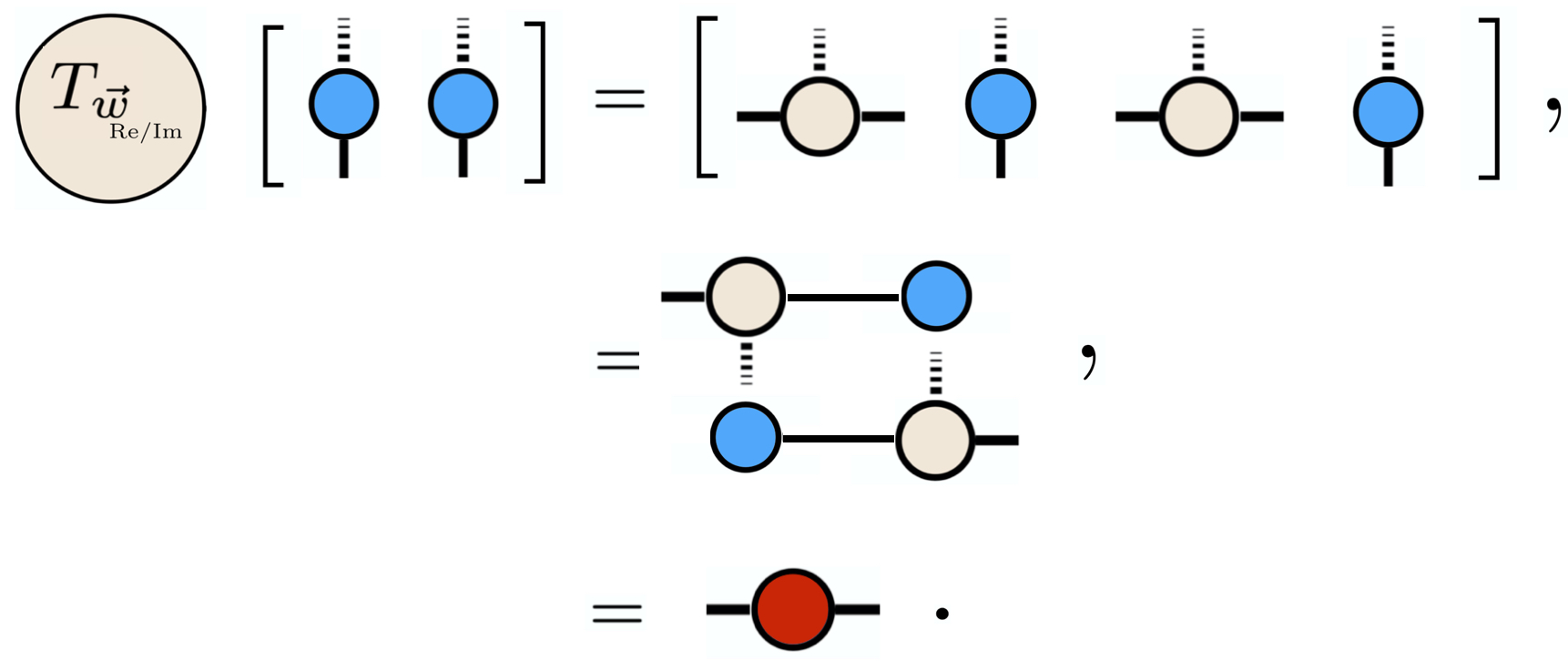} 
\caption{The feed-forward transformations $\mathbf{W}_\mathrm{Re/Im}$ used previously in \eref{perceptron_psi} are replaced by tensor-matrix contractions. The MPS $\vert\chi^{\mathrm{MPS}}_\theta \rangle\rangle$ is contracted with the MPO $T_{\vec{w}_\mathrm{Re/Im}}$ such that the final result is the full  $d\times d$ state.}
\label{fig_contraction_3}
\end{figure}

\begin{figure}[ht!]
\includegraphics[scale=0.15]{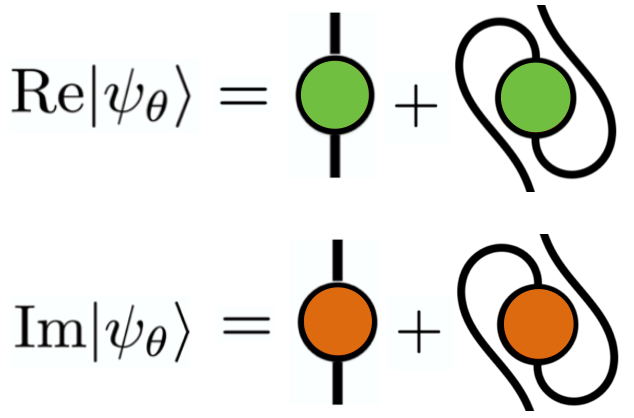} 
\caption{The real and imaginary parts of the wave function, before normalization and symmetrization. }
\label{fig_mps_sym}
\end{figure}

In order to explore some of the entanglement properties of the final state, let us recall the weakly entangled initial state
\begin{equation}\label{weak}
\vert \psi_i \rangle=|c_{\alpha_1},c_{\alpha_2}\rangle_{\mathrm{sym}}  \ \ \ \ \mathrm{rank}(\vert \psi_i \rangle)=2.
\end{equation}
Given this initial state, what will be the value of the rank of the final state $\vert\psi_\theta\rangle$ after random unitary evolution \eref{haar}. It is well known since the work of Nahum  \textit{et al.} \cite{nahum2017quantum,nahum2018operator} that the spread of entanglement entropy in random unitary circuits follows a very interestingly scaling. 
\begin{figure}[ht!]
\includegraphics[scale=0.145,left]{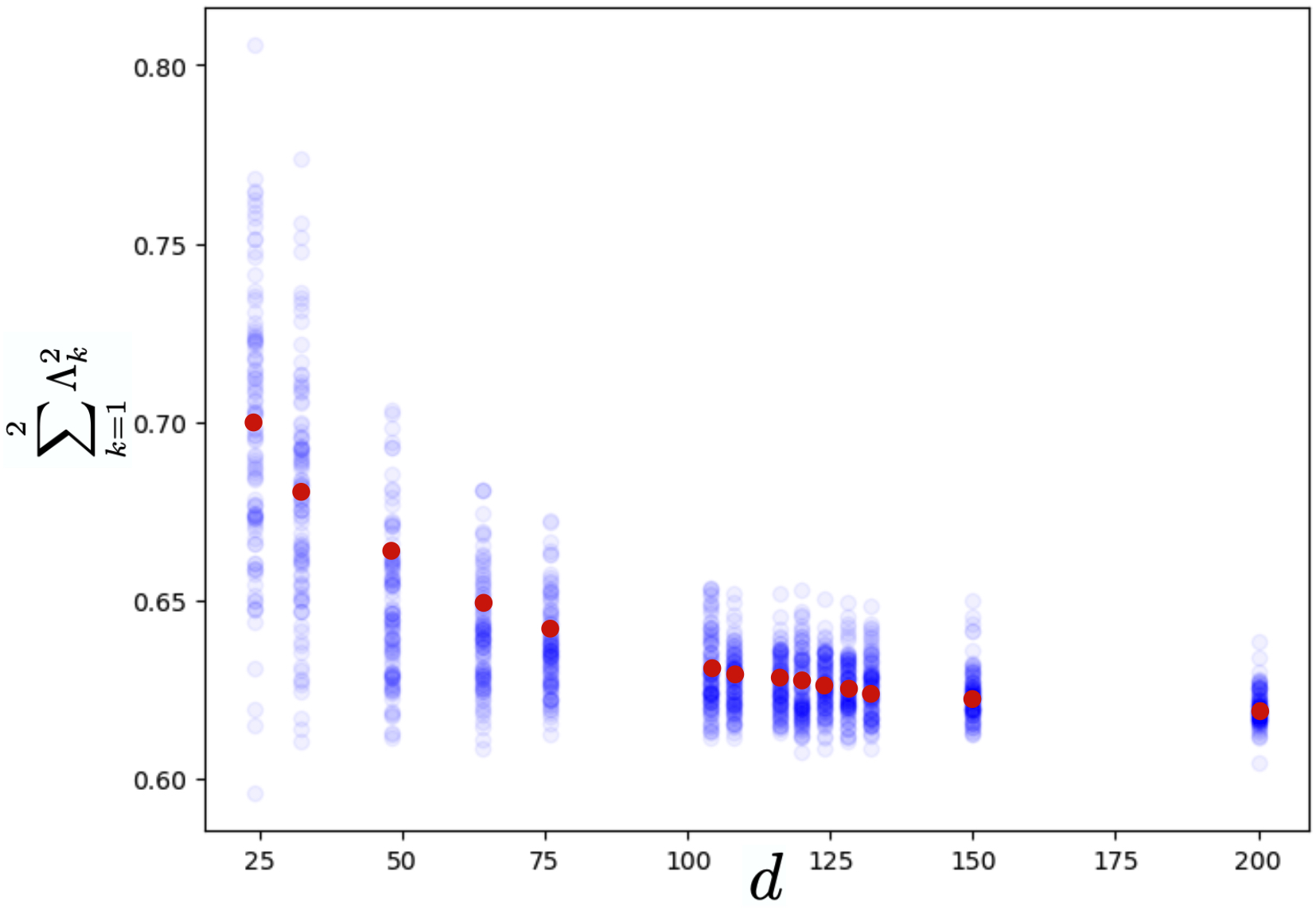} 
\caption{We show the value of $\sum_{k=1}^{2}\Lambda_k^{2}$ increasing $d$ and $1000$ random Haar realizations. The red dots represent mean values, they reach a finite value (around $62\%$) for large $d$, suggesting that some of the weakly entangled properties remain after random evolution.}
\label{fig_rank_2}
\end{figure}
The entanglement entropy grows linearly with time with super-diffusive fluctuations so that the system becomes maximally entangled at large time. 
Hopefully for a finite system and short time the spread of entanglement does not maximally entangled the entire Hilbert space. \efig{fig_rank_2} shows how well can the final state be described using the two leading components of the Schmidt decomposition, we see that large $d$, the system conserves a fair bit of its weakly entangled structure, as $\sum_{k=1}^{2}\Lambda_k^{2}\rightarrow 62\%$. 
\begin{figure}[ht!]
\includegraphics[scale=0.16]{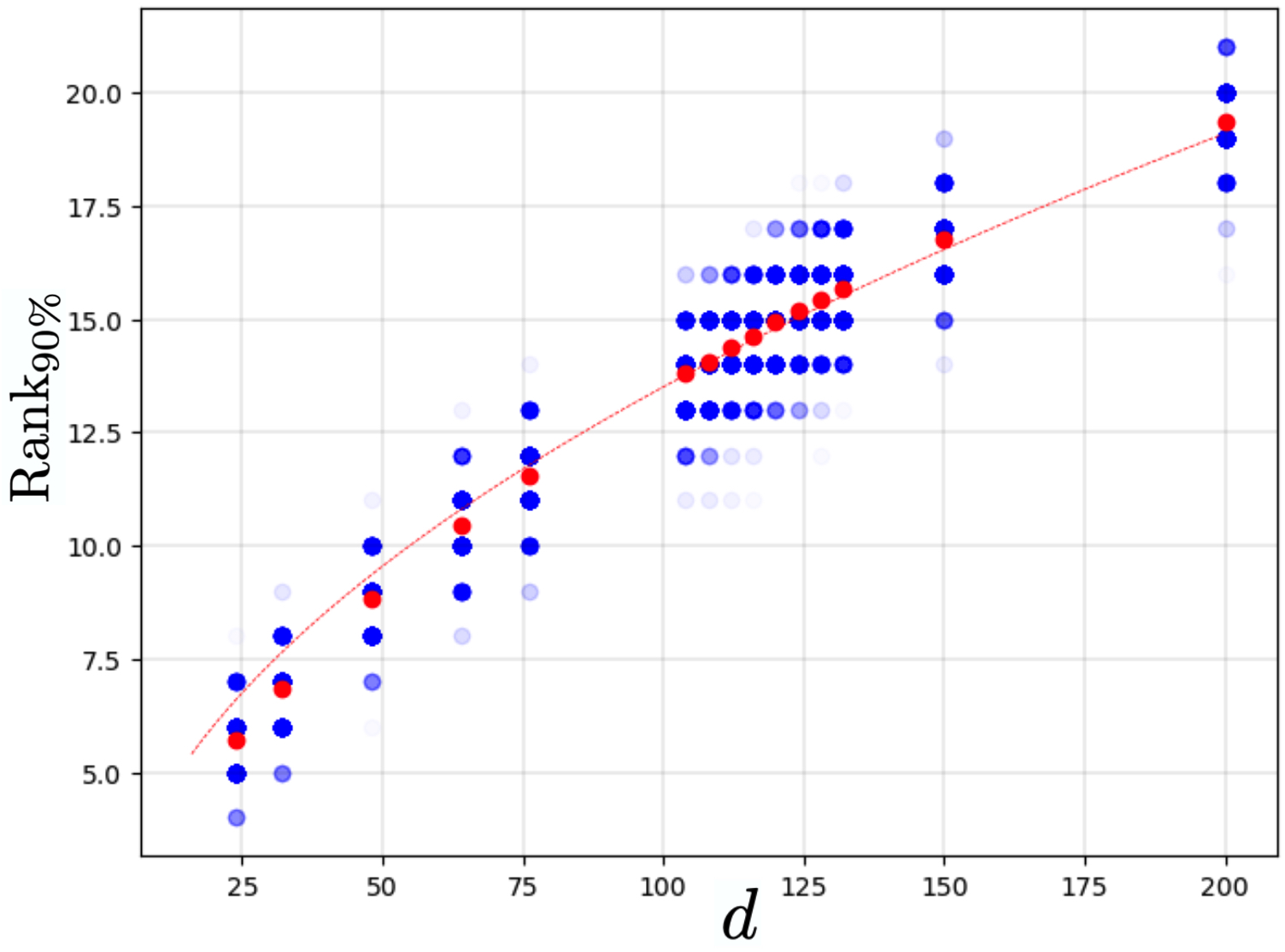} 
\caption{$\mathrm{Rank}_{90\%}$ is the rank of the Schmidt decomposition such that $\sum_{k=1}^{\mathrm{Rank}_{90\%}}\Lambda_k^{2}=0.9$. This rank is calculated for increasing $d$ for $1000$ Haar realizations. The red dots represent mean values and the red line is $\sqrt{d}$. For example for $d=64$ it requires around 10 Schmidt coefficients to sum up $90\%$ of the density matrix. This quantity will be used to choose a bond dimension for the tensor networks}
\label{fig_rank_90}
\end{figure}
On the other hand, \efig{fig_rank_90} shows how many components one needs to reach $90\%$ as $d$ grows larger. Contrary to a full rank initial state for which we need $0.9d$ components to reach $90\%$, here the state requires only $\sqrt{d}$ components. This rather slow scaling will be useful as a good approximation of the final state can be obtained with fewer components. 
The quantum state $|\psi_\theta\rangle$ of dimension $(d\times d)$ matrix can be written as a product of smaller $(d\times d_{b})$ matrices as showed in \efig{fig_mps_0}. It defines the simplest form of a 2-particle MPS with bond dimension $d_b$. For $d_b$ large enough, the relation becomes exact. In the case of a separable state $d_b\rightarrow 1$ and in the case of a maximally entangled state $d_b\rightarrow d$. We see that this approximation is useful when $d_b$ is reasonably small compared to $d$.
Usually in quantum many-body physics, the computation of matrix product state is performed using density matrix renormalization group of related variational algorithms that optimize directly the MPS to approximate the ground state of a local hamiltonian. In our particular situation, we adopt a different scheme and introduce a MPO (matrix product operator) \efig{fig_mpo} that once contracted with the MPS \efig{fig_mps_0} will hopefully approximate the quantum state. It is similar to the previous section where the quantum state was not directly parameterized, rather linear parameterized transformations were applied to a low-dimensional dummy state $|\chi_\theta\rangle$ \eref{perceptron} in order to fit the final state. The structure of low entanglement needs to be included in the MPO and MPS so that it will transpire in the final state.
So as to compare the computational power of tensor nets vs neural nets in a fair setting, we are simply substituting the two parameterized layers $T_{\vec{w}_0}$ and $T_{\vec{w}_\mathrm{Re/Im}}$ in \efig{fig_network} by low-dimensional tensor contractions. The second layer \eref{U1} which is $T_{\vec\theta}(\vec\theta)=(\mathrm{cos}(\vec\theta),\mathrm{sin}(\vec\theta))^{T}=\overset{\curvearrowleft}{\theta}$ remains the same as depicted in tensor diagrams \efig{fig_contraction_theta}.
The next layer takes in input the vector $\overset{\curvearrowleft}{\theta}$ and outputs a properly shaped matrix product state of dimension $(d\times d_{b},d\times d_{b})$  whereas the layers $T_{\vec{w}_\mathrm{Re/Im}}$ transform this state into real and imaginary parts of the target wave-function. The dotted lines represent the bond dimension $d_{b}$ of the matrix product state. The parameterized layers are replaced by tensor networks of respective dimensions $(d\times d_b \times n_{\mathrm{PS}},d\times d_b \times n_{\mathrm{PS}})$ and ($d\times d \times d_{b},d\times d \times d_{b}$) 
as depicted below in \efig{fig_contraction_0}. 
These operations can easily be performed by contraction of tensors as depicted in \efig{phase_contraction_1} and \efig{fig_contraction_3}. The number of parameters are $\#(T_{\vec{w}_\mathrm{Re}})=\#(T_{\vec{w}_\mathrm{Im}})=2d_{b}d^2$.
Compared to the previous section, where $T_{\vec{w_0}}$ would create a vector (\eref{perceptron_psi}) and $T_{\vec{w}_\mathrm{Re/Im}}$  would output a full $(d\times d)$ matrix (\eref{perceptron_psi}), now $T_{\vec{w_0}}$ creates a matrix product state and $T_{\vec{w}_\mathrm{Re/Im}}$ contract a tensor network with the MPS as shown in \efig{fig_contraction_3}.
The symmetry of the wave-function is simply enforced with the usual transpose trick show in \efig{fig_mps_sym}.
  \begin{figure}[ht!]
\includegraphics[scale=0.16]{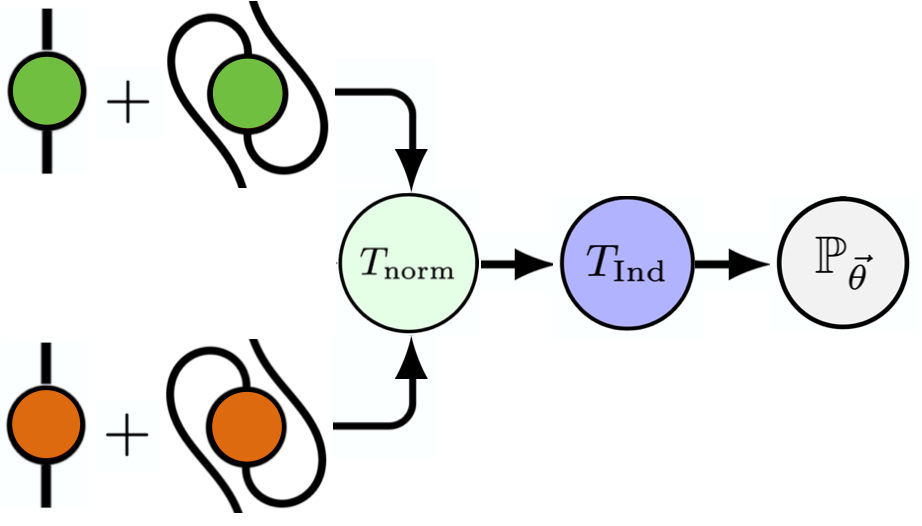} 
\caption{The real and imaginary symmetric matrices are fed to the rest of the network for proper normalization and indistinguishabilty.}
\label{fig_mps_net}
\end{figure}
The remaining of the architecture remains exactly the same, with the Boltzmann normalization \eref{boltzmann} and the triangularization \eref{triangle} transforming the state in a normalized indistinguishable bosonic probability distribution \efig{fig_mps_net}. The attentive reader will objet that the phase invariance as well as the symmetry properties could have been directly incorporated into the MPO \efig{fig_contraction_0} , still we have chosen to keep the architecture as similar as possible for comparison sake. In addition, MPS and other tensor networks, contrary to neural networks, are objects that are linear in nature, but here we keep the nonlinearities after the transformation $T_{\vec{w_0}}$ for the same reason. Let us mention that the only hyper-parameter in the network, beside $\beta$, is the bond dimension $b_d$ (similarly to the dimension $l$ for the neural net defined previously).
The training of the network and its histogram of errors are depicted in \efig{fig_training} where every other outside hyper-parameters (batch-size, epochs, learning rate...) are kept exactly the same.
\begin{figure}[ht!]
\includegraphics[scale=0.135]{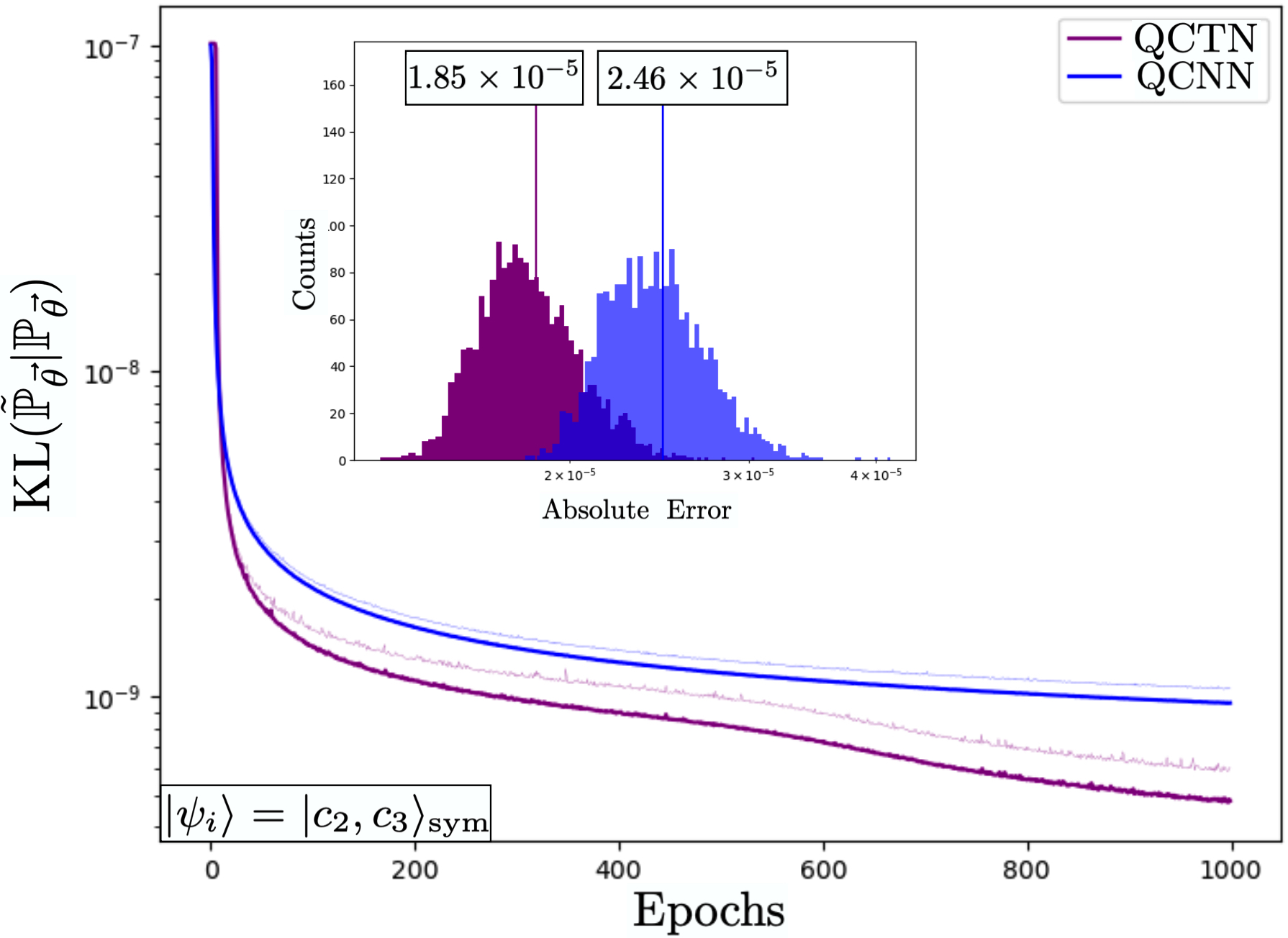} 
	\caption{Comparison of the quantum consistent neural (blue) and tensor (purple) nets. The plain color are the training losses whereas the shadowed lines are the corresponding validation losses. In the simulations, according to the finding of \efig{fig_rank_90}, we chose $d_b=10$ in the case $d=64$.}
\label{fig_training}
\end{figure}
  \begin{figure}[ht!]
\includegraphics[scale=0.21]{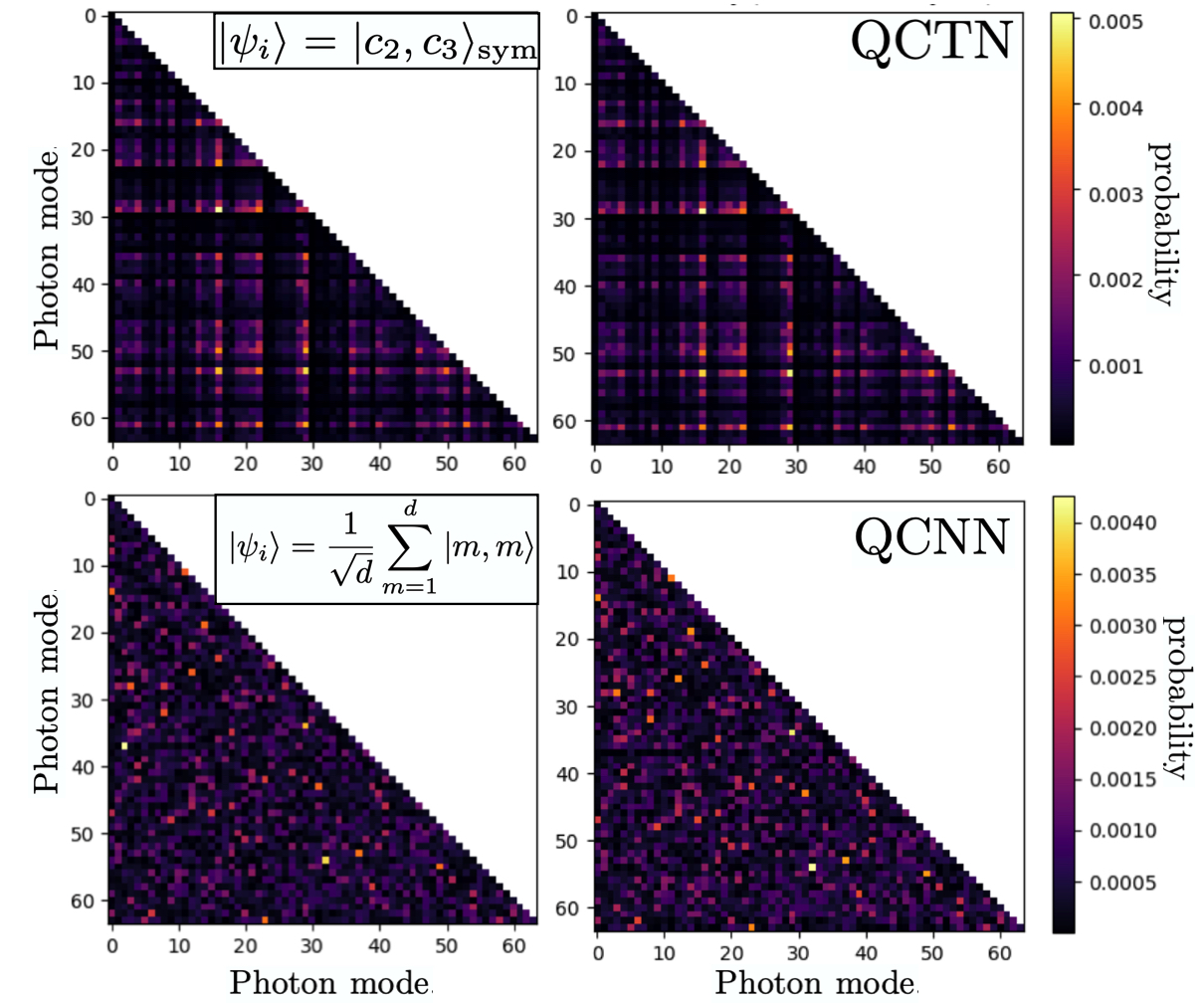} 
	\caption{Results of the two networks on the two initial states for the same $\vec\theta$ in the validation set. On the right side are the predictions and on the left the true distributions.}
\label{fig_comparaison}
\end{figure}
 When compared to the neural net, both the training and validation losses decrease more rapidly, demonstrating a better representation capacity. We can nonetheless notice a slight overfit of the tensor net (validation error bigger than training error), but not significantly enough to be a problem.  The histogram of errors of the tensor net on the validation set reveals a $33\%$ smaller mean absolute error than that of the neural network, which is a considerable improvement in precision.
The trainable parameters only live within the operators $T_{\vec{w_0}}$ and $T_{\vec{w}_\mathrm{Re/Im}}$. The number of parameters of the operators $T_{\vec{w}_\mathrm{Re/Im}}$ scales as $d^2$, similarly to the neural net (only because for $N=2$, $N\times d^2\sim d^N$) but now integrate the entanglement structure into the operations. Despite the same scaling, the tensor network possesses, thanks to the matrix product structure of its transformations, significantly fewer free parameters as we show in the table below. 
\begin{table}[ht!]
\centering
\begin{tabular}{ ||c | c | c| c| c| c || } 
\hline
 & $d=8$ & $d=16$ & $d=32$ & $d=64$ & $d=128$ \\
\hline
QC Neural Net & $14000$ &$52400$ & $206000$ & $820400$ & $3278000$ \\
QC Tensor Net & $4480$ &$14080$ & $48640$ & $179200$ & $686080$ \\
\hline
\end{tabular}
\caption{Number of trainable parameters of the $\mathrm{QCNN}$ vs the $\mathrm{QCTN}$ ($d_b=10$) for several $d$.}
\label{table}
\end{table}
As said before, the bond dimension appears to scale with the root ($d_b\sim d^{1/2}$) of the  dimension $d$ which makes the system not quite as efficient as if it was constant but still a lot better that what we would get with the maximally entangled state ($d_b\sim d$).
With this consideration in mind, we can confidently conclude that the tensor network outperforms the neural network at efficiently describing the full quantum evolution of the weakly entangled initial state \eref{weak}. Let us mention that, we are perhaps not using the full power of MPS's here, since we approximate a matrix $d\times d$ by a product of two smaller matrices $d\times d_b$. Another point of view would be to split the internal dimension $d$ into $\sqrt{d}$ small sub-systems of dimension $2$, and then construct a MPS of length $d$ with an MPO on top, to account for the entanglement between the $d$ modes instead of the global entanglement between the photons themselves. Nevertheless, the original approach is more than adequate for our purpose.

\begin{figure}[ht!]
\includegraphics[scale=0.12]{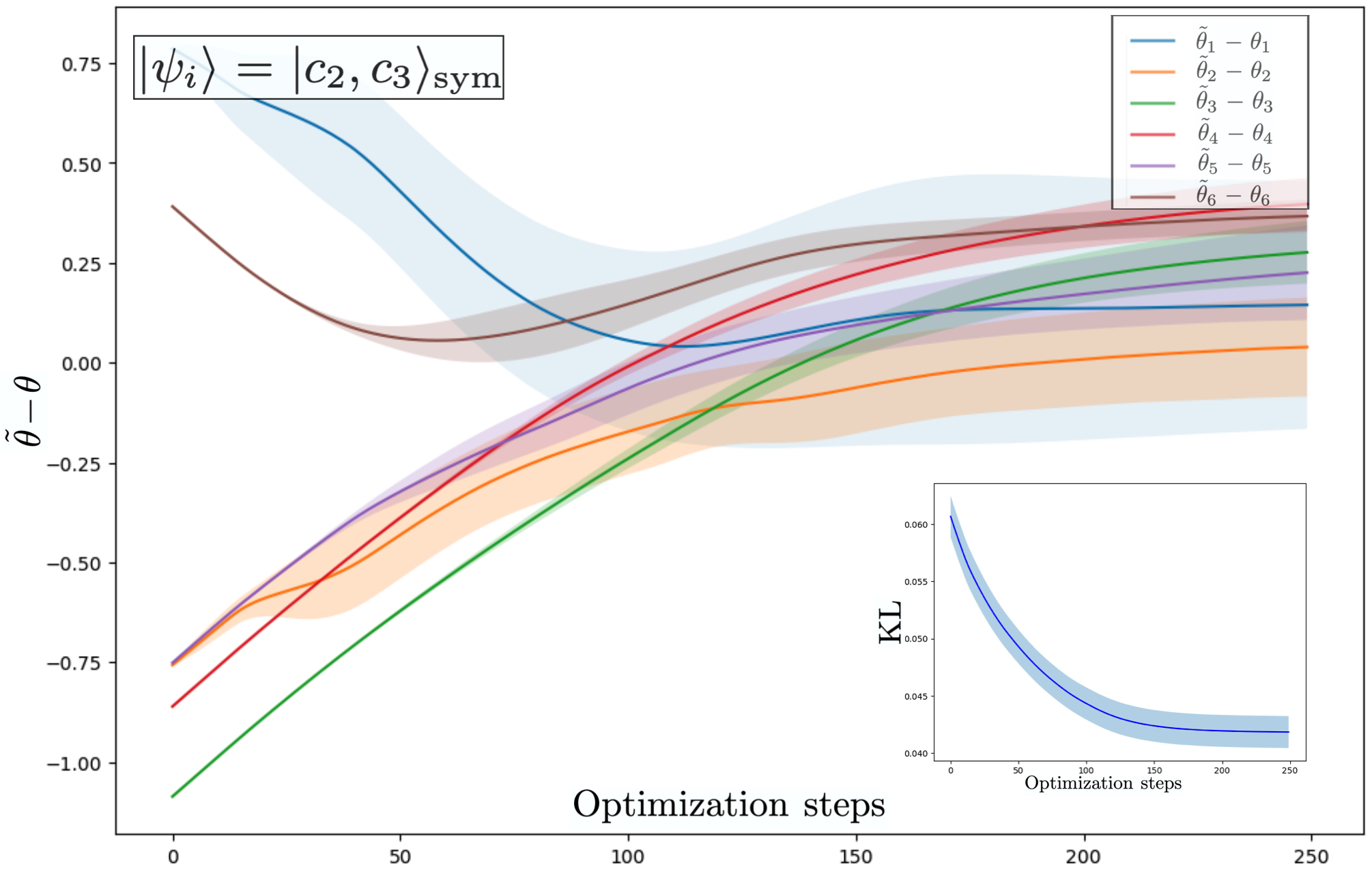} 
	\caption{Results of back-propagations of the tensor network for $d=64$ done on 100 distinct realizations of $\mathbb{P}_{\mathrm{emp}}(p=1000)$: the figure shows the convergence of the phase residuals and their standard deviation.}
\label{fig_tn_theta}
\end{figure}

Now that we have in our hands a trained network that is approximating the quantum probabilities accurately and efficiently, we can now move on to the back-propagation parameter optimization of said network, in the same fashion to the first section of this article. Let us approximate the probability matrix $\mathbb{P}_{\vec\theta}$ by the previously trained tensor net $\mathrm{TN}$ such that
\begin{equation}
\boxed{
\tilde{\mathbb{P}}_{\vec\theta}=\mathrm{TN}_{\vec{w}_{\bullet}}(\vec\theta),}
\end{equation}
where $\vec{w}_{\bullet}$ are the final weights of the trained network and $\vec\theta$ is the input phase vector. Thanks to the differentiability of the tensor-net (and the neural-net), the process of back-propagation can be performed in the exact same manner as we showed previously for the exact unitary evolution.
\begin{figure}[ht!]
\includegraphics[scale=0.15]{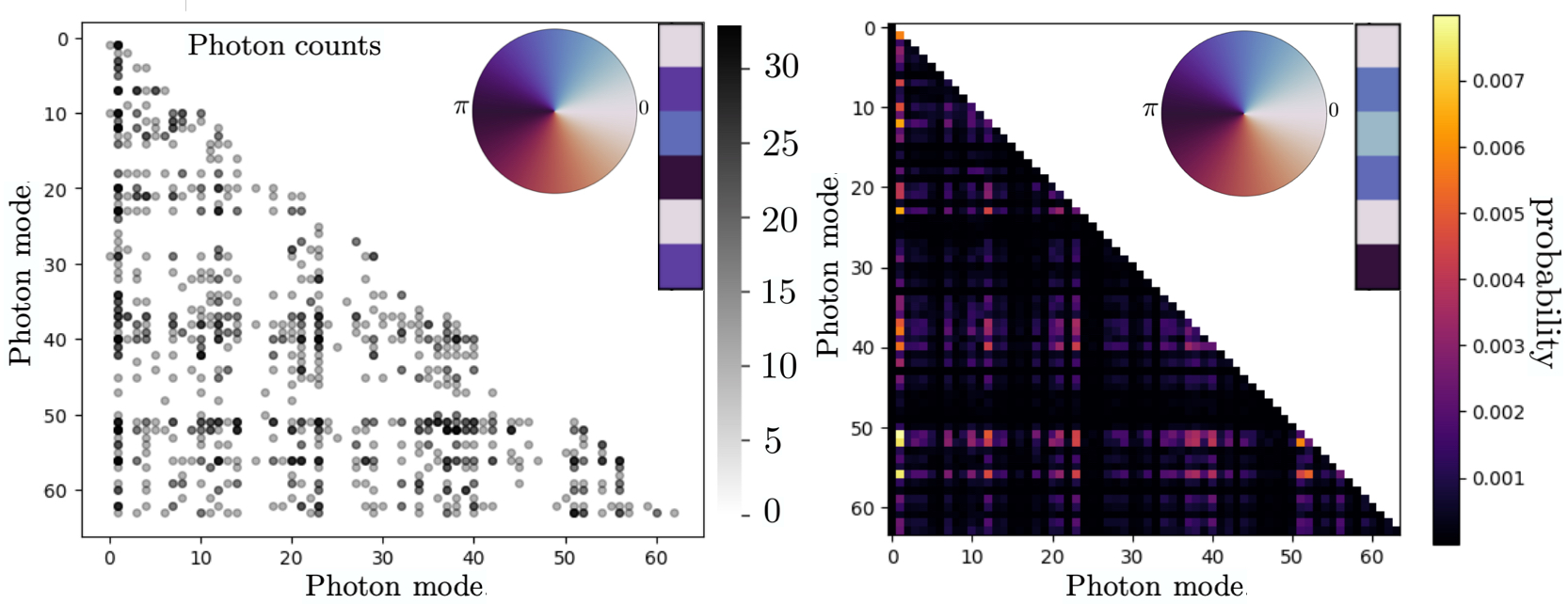} 
	\caption{Left: Photon counting measurement ($p=1000$) and the true phase-shift vector $\theta$. Right: The estimated vector $\tilde{\vec\theta}$, and its associated coincidence probability $\mathrm{TN}_{\vec{w}_{\bullet}}(\tilde{\vec\theta})$, computed with the optimization protocol described in the main text.}
\label{fig_tn_theta_1}
\end{figure}
 The optimization task remains the same as before, retrieve an estimated vector $\tilde{\vec\theta}$ such that $\mathrm{KL}(\mathrm{TN}_{\vec{w}_{\bullet}}(\tilde{\vec\theta})|\mathbb{P}_{\mathrm{emp}})$ is minimal. The process can be summarized as follow
 \begin{equation}\nonumber
\boxed{\mathrm{Find} \ \tilde{\vec\theta} \ \mathrm{such \ that} \ \mathrm{KL}\big(\mathrm{TN}_{\vec{w}_{\bullet}}(\tilde{\vec\theta}) |\mathbb{P}_\mathrm{emp}(p)\big) \ \mathrm{is \ min}.}
\end{equation}
 We use the same Adam optimizer \eref{adam} used in the first section of the article. Results of back-propagations of the tensor network for $d=64$ are shown in \efig{fig_tn_theta_1} and \efig{fig_tn_theta}. 
In these tests, the number of photons sent in the circuit ($p=1000$) remains the same making the ratio between the number of possible states and the number of samples much smaller (hence the poorer convergence of the estimation). Nonetheless, we observe a smooth convergence of the estimated vector of phase-shifts towards the true values. Knowing that the probing efficiency of the system is mostly dominated by the initial state and the number of samples, we argue that this convergence would be very similar to the one we would have obtained, had we used the exact unitary evolution instead (with a computer big enough) of the tensor net. Indeed having a tensor net only a few thousandth of percent less precise than the exact evolution would not impact significantly the precision of the quantum estimation. 

Let us mention that many other applications of optimizations and designs can be explored with the trained networks $\mathrm{NN}/\mathrm{TN}_{\vec{w}_{\bullet}}(\vec\theta)$. In particular, instead of trying to recover parameters of the circuit, we could optimize those parameters to hit a certain target of probability distribution. This could be helpful when trying to statistically create a specific quantum state by cloning or teleporting. Also the differentiability of the networks make computation of Fisher information and other derivative-based quantities easier in dimensions in which exact computation may be out of reach.

\section{More photons, boson sampling and conclusions}

Generalizing the idea to higher number of photons $N>2$ is difficult due to the non-determinantal nature of bosons. Indeed in the case of fermions, we know that the $N$-particles wave-function can be written as a determinant which can be computed efficiently in polynomial time on a computer. In contrast, the bosonic wave-function cannot be written as a determinant of a matrix but instead as a permanent which is extremely hard to compute on a classical computer. Propagation of indistinguishable photons through a random circuit such as \efig{fig_circuit} can be predicted using the knowledge of its transmission matrix $T$ (see \cite{tichy2014interference} for a nice review) which can be deduced form the unitary matrix. In particular, the probability
$\mathrm{Pr}(I \rightarrow F)$ of observing a certain configuration $F$ at the output assuming a certain input
state $I$ is directly given by calculating the permanent of a sub-matrix of the unitary matrix $U$. 
 For simplicity, let us consider the $N=3$, $d=4$ case of the initial state $\psi=|1,2,4\rangle_{\mathrm{sym}}$, with one photon in mode $1$, another one in mode $2$ and a third in mode $4$. Given the unitary matrix 
$U_{ij}$,
the coincidence probability that the detector clicks in mode $2$, $3$ and $4$ can be written as (we ignore the proportionality constant)
\begin{equation}
\mathrm{Pr}(2,3,4|1,2,4)\propto\Biggl|\mathrm{Per}\begin{pmatrix}
    U_{1,2} & U_{1,3}  & U_{1,4}  \\
    U_{2,2} & U_{2,3}  & U_{2,4} \\ 
    U_{4,2} & U_{4,3}  & U_{4,4} 
\end{pmatrix}.
\end{equation} 
Let us call this matrix $U^{124}_{234}$ where $U^{xyz}_{abc}$ means the sub-matrix constructed using columns $a$, $b$ and $c$ and rows $x$, $y$ and $z$. The permanent of a complex square matrix $U$ of size $d\times d$ is defined as
. The permanent of a complex square matrix $U$ of size $d\times d$ is defined as
\begin{equation}
\mathrm{Per} \ U=\sum_{\sigma\in S_d}\prod_{i=1}^{d}U_{i\sigma(i)},
\end{equation}
here the sum here extends over all elements $\sigma$ of the symmetric group $S_d$ i.e. over
all permutations of the numbers $1,2...d$. The entire probability distribution is simply the sum over all the amplitudes
\begin{eqnarray}\label{perm_eq}
&&\mathrm{Pr}(m_1,m_2,m_3|1,2,4)\propto\nonumber\\
&&\sum_{\substack{m_1,m_2,m_3=1 \\ m_1<m_2<m_3}}^{d=4}
\Biggl|\mathrm{Per}\begin{pmatrix}
    U_{1,m_1}  & U_{1,m_2} & U_{1,m_2}  \\
    U_{2,m_1}  & U_{2,m_2} & U_{2,m_3} \\ 
    U_{4,m_1}  & U_{4,m_2} & U_{4,m_3} 
\end{pmatrix}\Biggl|^{2},
\end{eqnarray}
\footnote{In the case $N=2$, $d=64$ that we focused on in the previous sections, it boils down to computing 2080 two by two permanent, which is actually a lot slower than doing the matrix computation (in a brute force way).}. For high number of photons $N>2$, the construction of the dataset as we described previously becomes hard to do. The matrix computation \eref{heisenberg} becomes more cumbersome to implement and long to compute, meanwhile the permanent formulation keeps its simplicity despite its exponentially-growing complexity. A realistic solution is to use the best known classical algorithm (the Ryser algorithm evaluates the permanent in $\mathcal{O}(2^{N-1}N^2)$) to sample randomly a fixed number of points in the distribution at each epochs and compare to the permanent formula \eref{perm_eq}.
\begin{figure}[ht!]
\includegraphics[scale=0.165]{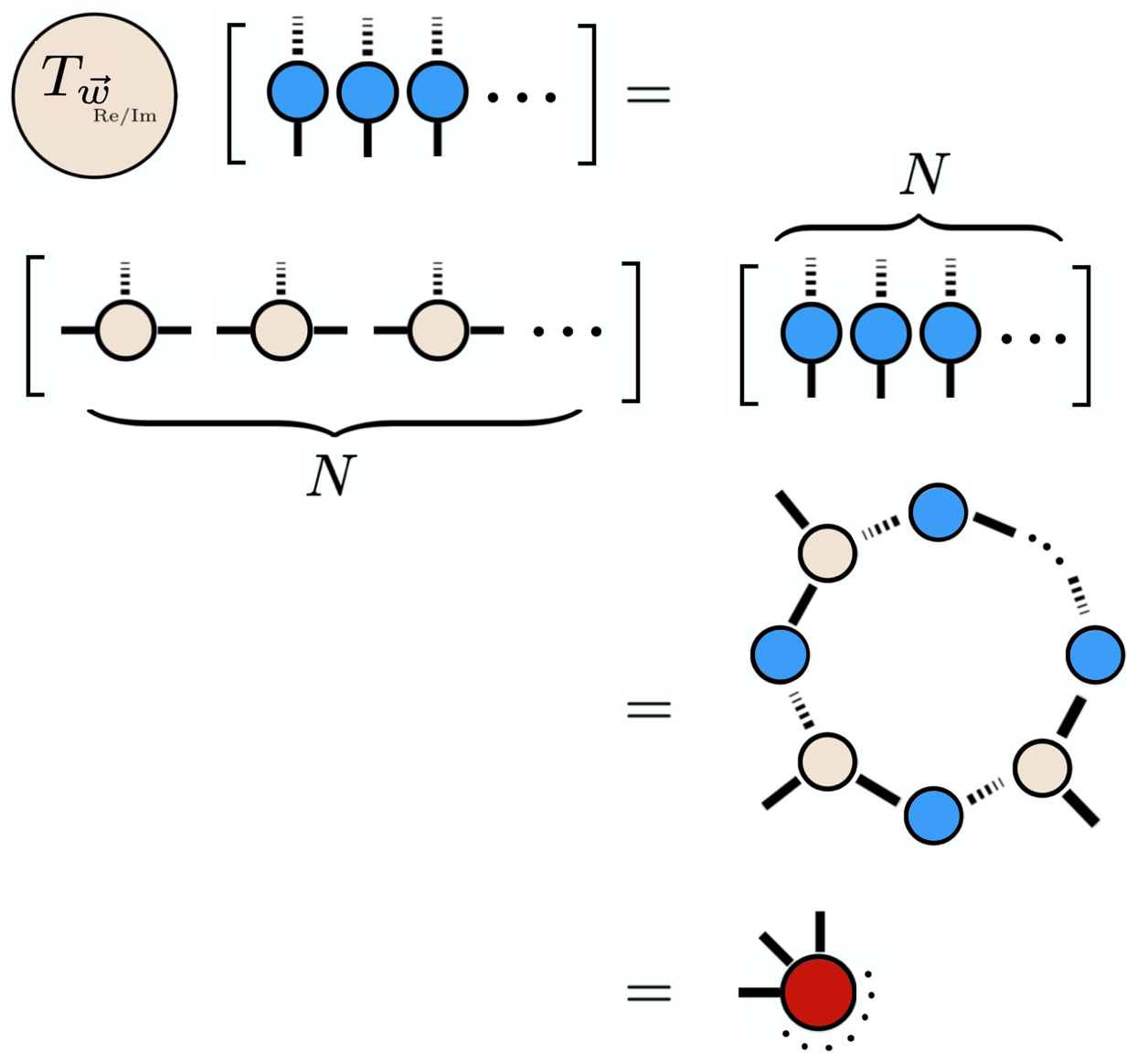} 
\caption{Extension of the MPO/MPS contraction for $N$ photons}
\label{fig_contraction_N}
\end{figure}

The $N=3$ loss function used to trained the network could take the following form ($C$ is a constant)
\begin{equation}\label{perm_loss}
\mathcal{L}_{\vec w}=\sum_{\vec\theta\in\mathcal{B}}\sum_{i,j,k \in R } \Big([\tilde{\mathbb{P}}_{\vec\theta}]_{ijk}- C\big|\mathrm{Per} \ U^{klm}_{ijk}\big|^{2} \Big)^{2},
\end{equation}
where $R$ is a set of randomly chosen coordinates in the space $(m_1,m_2,m_3)$, $klm$ represent the initial state, and $\mathcal{B}$ is the set of mini-batches and where $\big[\tilde{\mathbb{P}}_{\vec\theta}\big]_{ijk}=\mathrm{NN/TN}_{\vec{w}}(\vec\theta)$ is the parameterized neural/tensor net. The neural network defined in \efig{fig_network} is not an efficient approximation for $N>2$, the number of parameters scales as $d^N$ while the tensor net scales as $N\times d_b\times d^2$. The tensor net is an efficient approximation as long as $d_b$ does not grow exponentially with the number of photons, which is the case for weakly entangled states. In such cases, the tensor net defined previously, can be generalized without much modifications to higher number of photons such that $\big[\mathbb{P}_{\vec\theta}\big]_{ijk}=\mathrm{TN}_{\vec{w}}(\vec\theta)$, where $\mathrm{TN}_{\vec{w}}(\vec\theta)$ has a similar $N$-dimension generalization of the tensor contraction layers $T^{\mathrm{Re}/\mathrm{Im}}_{\vec{w}}$. This problem is related to the so-called boson sampling problem, which is a form of non-universal computer \cite{aaronson2011computational} that shows a demonstrated quantum advantage compared to classical sampling. Recently, the non-linear version of the boson sampling problem has been introduced \cite{spagnolo2023non} as a more expressive form of quantum computation. The neural and tensor networks presented in this work could be trained using a variation of \eref{perm_loss} and be used for classical quantum simulation purposes so to show whether or not these advantages are really genuine forms of quantum supremacy. Studies along those line have been recently published \cite{huang2019simulating,oh2021classical,oh2023tensor,liu2023simulating}.

In conclusion, we have introduced a deep learning approach to tackle the challenges associated with precise parameter estimation in entangled quantum optical systems in large dimensions. Our proposed method leverages quantum consistent neural/tensor networks to approximate the exact unitary evolution of strongly/weakly entangled states with high precision and efficiency. Through training the networks on quantum dynamics, we have demonstrated their effectiveness in enabling efficient parameter estimation in larger Hilbert spaces. This promising advancement opens new avenues for tackling quantum parameter estimation challenges, paving the way for enhanced understanding and manipulation of quantum states in complex quantum systems. The integration of neural and tensor networks provides a valuable tool for researchers in the pursuit of advancing quantum technologies and applications.

\textbf{Acknowledgements:} I greatly thank Christophe Chatelain for careful reading of the manuscript and all the contributors of the open-source libraries Tensorflow \cite{abadi2016tensorflow}, TensorNetwork \cite{roberts2019tensornetwork}, QuTip \cite{johansson2012qutip}, and Thewalrus \cite{gupt2019walrus}, thoroughly used for this work.

\bibliography{main}

\end{document}